\documentclass[a4paper, 12pt]{article}

\usepackage{amsfonts}
\usepackage{booktabs}
\usepackage{fullpage}
\usepackage[table]{xcolor} 
\usepackage{hyperref}
\usepackage{subcaption}
\usepackage{adjustbox}
\usepackage{algorithmicx}

\usepackage{acro}
\usepackage{amsmath}
\usepackage{amssymb}
\usepackage{amsthm}
\usepackage{mathtools}
\usepackage{graphicx}
\usepackage{cleveref}
\usepackage{upgreek}

\usepackage{subcaption}
\usepackage{nicefrac}

\usepackage{adjustbox}
\usepackage{algorithm}
\usepackage{algpseudocode}
\usepackage{natbib}
\usepackage{placeins}
\usepackage{physics}
\usepackage{wrapfig}

\newcommand{\E}{\mathop{\mathbb E}}

\newcommand{\R}{\mathbb{R}}

\newtheorem{proposition}{Proposition}

\DeclareMathOperator*{\argmin}{arg\,min}

\DeclareAcronym{mcmc}{short = {MCMC}, long  = {Markov chain Monte Carlo}}
\DeclareAcronym{rl}{short = {RL}, long  = {Reinforcement Learning}}
\DeclareAcronym{rlmh}{short = {RLMH}, long  = {Reinforcement Learning Metropolis--Hastings}}
\DeclareAcronym{mala}{short = {MALA}, long  = {Metropolis-adjusted Langevin algorithm}}
\DeclareAcronym{mdp}{short = {MDP}, long  = {Markov decision process}}
\DeclareAcronym{esjd}{short = {ESJD}, long  = {expected squared jump distance}}
\DeclareAcronym{lesjd}{short = {LESJD}, long  = {log ESJD}}
\DeclareAcronym{aar}{short = {AAR}, long  = {average acceptance rate}}
\DeclareAcronym{rmala}{short = {RMALA}, long  = {Riemannian MALA}}
\DeclareAcronym{ode}{short = {ODE}, long  = {ordinary differential equation}}
\DeclareAcronym{ddpg}{short = {DDPG}, long  = {deep deterministic policy gradient}}
\DeclareAcronym{cdlb}{short = {CDLB}, long  = {contrastive divergence lower bound}}
\DeclareAcronym{mmd}{short = {MMD}, long  = {maximum mean discrepancy}}
\DeclareAcronym{kl}{short = {KL}, long  = {Kullback--Leibler}}
\DeclareAcronym{td}{short = {TD}, long  = {temporal difference}}

\definecolor{ggplot1}{HTML}{E41A1C}
\definecolor{ggplot2}{HTML}{377EB8}

\begin{document}
\title{Harnessing the Power of Reinforcement Learning for Adaptive MCMC}
\author{Congye Wang$^{1,\star}$, Matthew A. Fisher$^{1,\star}$, Heishiro Kanagawa$^1$, \\ 
Wilson Chen$^2$, Chris. J. Oates$^{1,3}$ \\
\small $^1$Newcastle University, UK \\ 
\small $^2$University of Sydney, Australia \\
\small $^3$The Alan Turing Institute, UK \\
\small $^\star$equal contribution}
\maketitle

\begin{abstract}
Sampling algorithms drive probabilistic machine learning, and recent years have seen an explosion in the diversity of tools for this task.
However, the increasing sophistication of sampling algorithms is correlated with an increase in the tuning burden. 
There is now a greater need than ever to treat the tuning of samplers as a learning task in its own right.
In a conceptual breakthrough, \citet{wang2024reinforcement} formulated Metropolis--Hastings as a Markov decision process, opening up the possibility for adaptive tuning using \ac{rl}.
Their emphasis was on theoretical foundations; realising the practical benefit of \ac{rlmh} was left for subsequent work.
The purpose of this paper is twofold:
First, we observe the surprising result that natural choices of reward, such as the acceptance rate, or the expected squared jump distance, provide insufficient signal for training \ac{rlmh}.
Instead, we propose a novel reward based on the contrastive divergence, whose superior performance in the context of \ac{rlmh} is demonstrated.
Second, we explore the potential of \ac{rlmh} and present adaptive gradient-based samplers that balance flexibility of the Markov transition kernel with learnability of the associated \ac{rl} task.
A comprehensive simulation study using the \texttt{posteriordb} benchmark supports the practical effectiveness of \ac{rlmh}.
\end{abstract}

\section{Introduction}

Probabilistic machine learning leverages probability theory as the basis for making inferences and predictions in a machine learning task.
Although only a subset of machine learning, the probabilistic approach is particularly useful in applications where it is important to quantify uncertainty, handle incomplete data, or incorporate prior knowledge into a model. 
At the core of probability theory is the concept of conditioning, and we are typically interested in the distribution of unobserved quantities conditional on an observed training dataset.
Such conditional distributions are characterised by Bayes' theorem, and are called \emph{posterior} distributions in the Bayesian context.
However, the lack of a closed form for the normalisation constant in Bayes' theorem necessitates numerical approximation in general.

Since the early development of posterior sampling methodology, such as \ac{mcmc}, the scope and ambition of numerical methods for posterior approximation has expanded to include techniques such as variational approximations \citep{fox2012tutorial}, sequential Monte Carlo \citep{chopin2020introduction}, gradient flows \citep{chen2023sampling}, normalising flows \citep{rezende2015variational}, and more recently samplers based on diffusion models \citep{blessing2025underdamped}.
As a result, one can argue that the main research challenge is no longer the design of new methods for posterior approximation, but rather the \emph{tuning} of existing methods to perform well on a given task.
Unfortunately, the tuning problem is difficult; despite half a century of research, the design and use of convergence diagnostics for tuning \ac{mcmc} remains an active research topic. 
Further, the increasing sophistication of emerging posterior approximation methods is usually associated with increasing difficulty of the associated tuning task.
These observations lead us to contend that there is now a greater need than ever to treat the tuning of samplers as a learning task in its own right.

The idea in this manuscript is to harness the potential of \acf{rl} to provide a powerful and principled solution to the tuning of \ac{mcmc}. 
Our inspiration comes from \citet{wang2024reinforcement}, who demonstrated how Metropolis--Hastings samplers can be formulated as Markov decision processes, and are therefore amenable to tuning using modern \ac{rl}.
That framework was dubbed \textit{Reinforcement Learning Metropolis--Hastings} (RLMH), and its correctness (i.e. ergodicity) was established using appropriately modified techniques from the literature on adaptive \ac{mcmc}.
However, actually realising the practical benefit of \ac{rlmh} was left for subsequent work.
The purpose of this paper is twofold:
First, we observe the surprising result that natural choices of reward, such as the acceptance rate, or the expected squared jump distance, provide insufficient signal for training \ac{rlmh}.
Instead, we propose a novel reward based on the contrastive divergence, whose superior performance in the context of \ac{rlmh} is demonstrated.
Second, we explore the potential of \ac{rlmh} and present adaptive gradient-based samplers that balance flexibility of the Markov transition kernel with learnability of the associated \ac{rl} task.
A comprehensive simulation study using the \texttt{PosteriorDB} benchmark supports the effectiveness of \ac{rlmh}.

\subsection{Related Work}

Throughout we consider a posterior distribution $P$ with positive and differentiable density $p(\cdot)$ supported on $\mathbb{R}^d$, but in principle much of our discussion can be generalised with additional technical effort.

\paragraph{Gradient-Based MCMC}

There are an enormous diversity of \ac{mcmc} algorithms in the literature, but for high-dimensional posterior distributions it is usually essential to have access to gradients of the target.
The most common mechanism through which gradient information is exploited is through the \emph{overdamped Langevin diffusion}
\begin{align}
    \mathrm{d}X_t = (\nabla \log p)(X_t) \mathrm{d}t + \sqrt{2} \mathrm{d}W_t \label{eq: sde}
\end{align}
which, under regularity conditions, defines a stochastic process $(X_t)_{t \geq 0}$ whose law converges to $P$ in the $t \rightarrow \infty$ limit \citep[see e.g.][]{eberle2016reflection}.
Since it is usually not possible to exactly simulate \eqref{eq: sde}, one can consider a (biased) Euler--Maruyama discretisation
\begin{align}
    X_{n+1}^* = X_n + \epsilon (\nabla \log p)(X_n) + \sqrt{2 \epsilon} Z_n , \qquad Z_n \stackrel{\text{iid}}{\sim} N(0,1) , \label{eq: mala}
\end{align}
as a proposal, which can then be either accepted or rejected within the Metropolis--Hastings framework to ensure convergence to the correct target; this is known as the \ac{mala} \citep{robert1996exponential}.
Selection of the \emph{step size} $\epsilon > 0$ is critical; for $\epsilon$ too small, many iterations will be required, while for $\epsilon$ too large the discretisation error in \eqref{eq: mala} will be substantial and the proposed states will be rejected.
Unfortunately, the range of reasonable values for $\epsilon$ is often extremely small \citep{livingstone2022barker}.
This observation motivated the development of \ac{rmala}, 
\begin{align}
    X_{n+1}^* = X_n + G(X_n)^{-1} (\nabla \log p)(X_n) + \sqrt{2} G(X_n)^{-1/2} Z_n , \qquad Z_n \sim N(0,1) ,   \label{eq: rmala}
\end{align}
where $G(X_n)$ acts as a position-dependent preconditioning matrix.
Some possible choices for $G(\cdot)$ based on second derivatives of the target were proposed in \citet{girolami2011riemann}, and sufficient conditions for ergodicity were established in \citet{roy2023convergence}.
Second derivative information enables the step size to be scaled in a manner commensurate with the curvature of the target, and \ac{rmala} can be an effective sampling method.
Unfortunately, the computational cost associated with obtaining higher order derivatives is often substantial\footnote{For instance, in the setting of parameter inference for \acp{ode}, to obtain first order gradients the \ac{ode} system can be augmented with an additional $O(d)$ variables, called \emph{sensitivities}, and this augmented system of \acp{ode} can be integrated forward.  For second order derivatives, an additional $O(d^2)$ variables must be included.  Unfortunately numerical integration of large systems of \acp{ode} can be a challenging task, and fine time steps can be required to properly resolve the dynamics of all the additional variables that are included.}, and as a result \ac{rmala} is not widely used.
Compromises include taking $G(\cdot) = \epsilon^{-1} G_0$ to be an appropriately scaled constant matrix \citep{roberts2002langevin}, or $G(\cdot) = \epsilon^{-1} D_0$ where $D_0$ is a diagonal matrix and the state space of the Markov chain is augmented to include $\epsilon$ as a variable that takes values in a discrete set \citep{biron2023automala,liu2024autostep}.

\paragraph{Adaptive MCMC}

Due to the challenge of tuning \ac{mcmc}, on-line (or \emph{adaptive}) tuning methods have considerable practical appeal \citep{haario2006dram,roberts2009examples}.
Several creative solutions have been proposed, for instance synthesising ideas from Bayesian optimisation \citep{mahendran2012adaptive}, divergence minimisation \citep{coullon2023efficient,kim2025tuning}, and generative diffusion models \citep{hunt2024accelerating}.
However, altering the Markov chain transition probabilities based on the current sample path breaks the Markov dependency structure and ergodicity can be lost.
Theoreticians have sought to discover sufficient criteria under which adaptive \ac{mcmc} methods can asymptotically sample from the correct target \citep{atchade2005adaptive,andrieu2006ergodicity,atchade2011adaptive}.
The most well-known of these are \emph{containment} and \emph{diminishing adaptation}; the first ensures the Markov chain transition probabilities do not become degenerate, and the second ensures the transition probabilities converge to a limit \citep{roberts2007coupling}.
Focussing on Metropolis--Hastings, adaptive algorithms attempt to optimise an explicit performance criterion by iteratively refining parameters of the proposal.
Several criteria are available, but the most common\footnote{Some less widely used criteria include the asymptotic variance of quantities of interest \citep{andrieu2001controlled}, the raw acceptance probabilities \citep{titsias2019gradient}, and divergences between the proposal and the target distributions \citep{andrieu2006ergodicity,dharamshi2023sampling}.} choices are the \ac{aar} and the \ac{esjd} \citep{pasarica2010adaptively}.
Theoretical analysis provides guidance on which \ac{aar} to target\footnote{An optimal \ac{aar} of 0.234 was established for a class of random walk Metropolis--Hastings algorithms in \citet{gelman1997weak}, while the value 0.574 was established for \ac{mala} in \citet{roberts1998optimal}.}, but since this is limited to idealised settings its practical relevance is debated \citep{potter20150}.
For \ac{esjd}, higher values are preferred, but a high \ac{esjd} does not itself guarantee that the posterior is well-approximated (c.f. \Cref{subsec: inadequacy}). 
For a relatively inflexible parametric proposal, these existing criteria can often assist in the identification of suitable parameter values, but for more flexible classes of proposal the information provided by the \ac{aar} and \ac{esjd} can be insufficient to identify a suitable proposal.

\paragraph{RL Meets Adaptive MCMC}

Online decision problems with a Markovian structure are the focus of \ac{rl}; it therefore appears natural to consider using \ac{rl} for adaptive \ac{mcmc}.
Recall that a \ac{mdp} consists of a \textit{state} set $\mathcal{S}$, an \textit{action} set $\mathcal{A}$, and an \textit{environment} which determines both how the state $s_n$ is updated to $s_{n+1}$ in response to an action $a_n$, and the (scalar) \emph{reward} $r_n$ that resulted.
\ac{rl} refers to a broad class of methods that attempt to learn a \emph{policy}\footnote{Throughout this paper we focus on \emph{deterministic} policies, though stochastic policies are also widely used in \ac{rl}.  A deterministic policy is essential to \ac{rlmh}, as additional stochasticity in the policy can introduce render the Markov chain no longer $P$-invariant.} $\pi : \mathcal{S} \rightarrow \mathcal{A}$, meaning a mechanism to select actions $a_n = \pi(s_n)$, which aims to maximise an expected cumulative (discounted) reward \citep{kaelbling1996reinforcement}.
A natural strategy is to define the state $s_t$ of the \ac{mdp} to be the state $X_t$ of the Markov chain, and the action to be parameters of the Markov transition kernel.
This approach was called \textit{Reinforcement Learning Accelerated MCMC} in \cite{chung2020multi}, where a finite set of Markov transition kernels were developed for a particular multi-scale inverse problem, and called \textit{Reinforcement Learning-Aided MCMC} in \cite{wang2021reinforcement}, where \ac{rl} was used to learn frequencies for updating the blocks of a Gibbs sampler.
The success of these methods relied on limiting the number of transition kernels that were considered; for a flexible class of transition kernels the action would become high-dimensional, a regime to which \ac{rl} is not well-suited.
A breakthrough occurred in \citet{wang2024reinforcement}, where it was shown how Metropolis--Hastings algorithms can be cast as \acp{mdp} via an augmented state $s_n = [X_n,X_{n+1}^*]$, consisting of the current and proposed state of the Markov chain, and an action $a_n = [\phi(X_n),\phi(X_{n+1}^*)]$.
The augmented state is necessary because the Metropolis--Hastings acceptance probability requires both the parameters $\phi(X_n)$ of the forward proposal and the parameters $\phi(X_{n+1}^*)$ of the reverse proposals to be provided within the \ac{mdp}.
The key insight of \ac{rlmh}, in contrast to earlier work, is to parametrise the (local) proposal distribution instead of the (global) transition kernel, simultaneously enabling flexible transition kernels while keeping actions low-dimensional.   
The authors of that work focussed on theoretical analysis, establishing sufficient conditions on \ac{rlmh} for containment and diminishing adaptation to hold, while their methodological development was limited to a gradient-free random walk Metropolis--Hastings algorithm in which $X_{n+1}^*$ was sampled from a distribution with mean $\phi(X_n)$ and the function $\phi(\cdot)$ was learned using \ac{rl}.
Realising the full potential of \ac{rlmh}, including for gradient-based algorithms, was left for future work.

\subsection{Outline of the Paper}

The aim of this paper is to develop efficient gradient-based adaptive \ac{mcmc} algorithms using the framework of \ac{rlmh}, which we recall in \Cref{subsec: rl}.
Motivated by a trade-off between flexibility of the transition kernel and learnability of the associated \ac{rl} task (i.e. seeking a low-dimensional action space), we focus on \ac{rmala} \eqref{eq: rmala} with $G(\cdot) = \epsilon(\cdot)^{-1} G_0$ where $G_0$ is a fixed symmetric positive definite matrix and $\epsilon(\cdot)$ is a positive \emph{step-size function} to be learned using \ac{rl}.
Intuitively, such a position-dependent preconditioner $G$ has the potential to overcome the tuning difficulties of \ac{mala}, while controlling the difficulty of the \ac{rl} task and not requiring second order derivative information on the target.
A key technical difficulty is that the usual performance criteria of \ac{aar} and \ac{esjd} are inadequate for \ac{rlmh}; we explain the issue in \Cref{subsec: inadequacy} and propose a novel alternative criterion based on contrastive divergence in \Cref{subsec: cd}, which may be of independent interest.
A comprehensive empirical assessment based on the \texttt{posteriordb} benchmark is presented in \Cref{sec: empirical} and a discussion is provided in \Cref{sec: discuss}.

\section{Methods}

The notation and set-up for \ac{rlmh} are recalled in \Cref{subsec: rl}.
The inadequacy of standard performance criteria is explained in \Cref{subsec: inadequacy}, and our novel performance criterion, based on contrastive divergence, is presented in \Cref{subsec: cd}.
Though our main focus is \ac{mala} and \ac{rmala}, we briefly discuss opportunities to leverage \ac{rl} also for other \ac{mcmc} algorithms in \Cref{ap: Barker}.

\subsection{Reinforcement Learning for RMALA}
\label{subsec: rl}

This section explains how \ac{rl} can be employed to tune \ac{rmala} using the framework of \ac{rlmh}.
Our setting is \eqref{eq: rmala} with the specific choice $G(\cdot) = \epsilon(\cdot)^{-1} G_0$, where $\epsilon : \mathbb{R}^d \rightarrow (0,\infty)$ is to be learned using \ac{rl}.
Since we seek flexible deterministic policies $\epsilon(\cdot)$ for continuous action spaces, we employed a \ac{ddpg} \citep{lillicrap2015continuous} method\footnote{For concreteness we present the methodology using \ac{ddpg} in the main text, but we note that other \ac{rl} algorithms compatible with deterministic policies and continuous action spaces could be considered, such as the TD3 method of \citet{fujimoto2018addressing}.}. 
That is, the function $\epsilon \equiv \epsilon_\theta$ is taken to be a deep neural network with parameters denoted $\theta \in \mathbb{R}^p$.
Since we will be learning $\theta$ in an online manner, we write $\theta_n$ to denote the value of the parameters at iteration $n$.
Both the initial state $X_0$ of the Markov chain and the initial parameters $\theta_0$ are randomly initialised.
The resulting algorithm is fairly straight-forward and consists of alternating, with some fixed frequency, between the following two steps:

\paragraph{Update $X_n$ via RMALA}
Given the current state $X_n$ of the Markov chain and the current parameters $\theta_n$ of \ac{rmala}, propose a new state
\begin{align*}
    X_{n+1}^* | X_n \sim Q_{\theta_n}(\cdot | X_n) := N( X_n + \epsilon_{\theta_n}(X_n) G_0^{-1} (\nabla \log p)(X_n) , 2 \epsilon_{\theta_n}(X_n) G_0^{-1} ) .
\end{align*}
Letting $q_{\theta_n}(\cdot | x)$ denote a density for $Q_{\theta_n}(\cdot | x)$, we then set $X_{n+1}$ equal to $X_{n+1}^*$ with probability
\begin{align*}
    \alpha(X_{n+1}^* | X_n) := \min\left\{ 1 , \frac{p(X_{n+1}^*)}{p(X_n)} \frac{q_{\theta_n}(X_n|X_{n+1}^*)}{q_{\theta_n}(X_{n+1}^* | X_n)} \right\}
\end{align*}
else set $X_{n+1}$ equal to $X_n$.

\paragraph{Update $\theta_n$ via Policy Gradient}
Conceptually, a \emph{policy gradient} method attempts to take a step of gradient ascent 
\begin{align}
    \theta_{n+1} = \theta_n + \beta_n (\nabla J)(\theta_n), \qquad 
    J(\theta) = \mathbb{E}\left[ \sum_{n = 1}^\infty \gamma^{n-1} r_n(s_n,a_n,s_{n+1}) \right] , \label{eq: J def}
\end{align}
where $\beta_n \in (0,\infty)$ is a learning rate, $\gamma \in (0,1)$ is a discount factor\footnote{If the Markov chain is initialised at stationarity, i.e. $X_0 \sim P$, then the rewards $r_n$ are identically distributed and $J(\theta) = \frac{\gamma}{1 - \gamma} \mathbb{E}[r_1(s_1,a_1,s_2)]$, meaning the discount factor $\gamma$ can be absorbed into the learning rate, or simply ignored.}, and $r_n$ is a reward, to be specified.
Thus, for \ac{rlmh}, the reward $r_n$ can (only) depend on the current state $s_n = [X_n,X_{n+1}^*]$, action $a_n = [\epsilon_{\theta_n}(X_n),\epsilon_{\theta_n}(X_{n+1}^*)]$, and subsequent state $s_{n+1} = [X_{n+1},X_{n+1}^*]$ of the \ac{mdp}.
Of course, the objective $J$ is intractable in general, and most of the research effort into policy gradient methods is invested in developing approximations to the policy gradient.
\ac{ddpg} achieves this by training a differentiable \emph{critic} network to predict the \emph{value} associated to a state-action pair $(s,a) \in \mathcal{S} \times \mathcal{A}$; since we use \ac{ddpg} off-the-shelf, we refer the reader to \citep{lillicrap2015continuous} for full detail.

\medskip

Implemented efficiently, the runtime of \ac{rlmh} is $O(n)$ and the storage requirement is $O(1)$, as for general \ac{mcmc}.
Despite the apparent complexity of the $\theta_n$ update via policy gradient, general sufficient conditions for ergodicity of \ac{rlmh} were established in \citet{wang2024reinforcement}.
In brief, containment can be enforced through parametrisation of the policy, so that the transition kernel is uniformly ergodic when the parameters $\theta$ are constrained to a compact set, while diminishing adaptation can be enforced by employing a summable learning rate $(\beta_n)_{n \in \mathbb{N}}$, in conjunction with gradient clipping to ensure that the sequence $(\theta_n)_{n \in \mathbb{N}}$ remains confined to a bounded set.  
Fortunately, both a summable learning rate and gradient clipping are relatively standard in the context of \ac{ddpg}.
The technical analysis of uniform ergodicity is much more challenging for gradient-based samplers compared to the gradient-free case studied in \citet{wang2024reinforcement}, and we do not attempt it in this work.
Instead, we simply note that one can in practice set the learning rate $\beta_n$ to zero after a fixed number of iterations have been performed, so that ergodicity is immediate from ergodicity of \ac{rmala} \citep{roy2023convergence}.

\subsection{Inadequacy of `Natural' Rewards}
\label{subsec: inadequacy}

Designing an informative reward is critical to the success of \ac{rl}.
Recall that, in the framework of \ac{rlmh}, the reward $r_n$ may depend on the current state $s_n = [X_n, X_{n+1}^*]$, the action $a_n = [\epsilon_{\theta_n}(X_n), \epsilon_{\theta_n}(X_{n+1}^*)]$, and the subsequent state $s_{n+1} = [X_{n+1}, X_{n+1}^*]$.
This rules out the \ac{aar}, which requires knowledge of the full sample path, but permits the apparently natural choice of the squared jump distance $r_n = \|X_n - X_{n+1}\|^2$, motivated by the theoretical connection between \ac{esjd} and mixing times \citep{sherlock2009optimal}, or indeed a Rao--Blackwellised version $r_n = \alpha_n \|X_n - X_{n+1}^*\|^2$ where the randomness due to the accept/reject step is integrated out.
However, both choices provide inadequate signal for \ac{rlmh}; the policy gradient is essentially flat when most samples are rejected.
A possible remedy is to increase the dynamic range; to this end, the \ac{lesjd} reward $r_n = \log(\alpha_n) + 2 \log \|X_n - X_{n+1}^*\|$ was proposed in \citep{wang2024reinforcement}.
Nevertheless, \ac{esjd} and related rewards such as \ac{lesjd} depend critically upon stationarity of the Markov chain; consider sampling from a standard Gaussian using a Metropolis--Hastings proposal $X_{n+1}^* \sim N(- X_n , \epsilon )$.
If the Markov chain is far in the tail, e.g. $X_0 = -100$, then optimising $\epsilon$ via policy gradient ascent targeting \ac{esjd} leads to $\epsilon = 0$, for which large jumps between $\pm 100$ are always accepted but ergodicity is lost.
A technical contribution of this work is to propose a more suitable reward that respects the structure of the \ac{mdp} and improves stability of training \ac{rlmh} by avoiding the pathologies of \ac{esjd} and \ac{lesjd}.

\subsection{Contrastive Divergence Lower Bound}
\label{subsec: cd}

A natural performance criterion for Metropolis--Hastings would be the \ac{kl} divergence between proposal and target, but this requires a sufficiently tractable proposal and a method through which the \ac{kl} can be approximated \citep{arbel2021annealed,kim2025tuning}. 
Instead we seek a general criterion, that is similar in spirit to \ac{kl} but can be easily estimated using only information available in the \ac{mdp} formulation of \ac{rlmh}. 

To begin, let $P_n$ denote the law of $X_n$ and let $p_n$ be a density for $P_n$, assumed to exist.
The convergence of $P_n$ to $P$ can be quantified using the \ac{kl} divergence $D_{\mathrm{KL}}(P_n || P) := \int \log( \mathrm{d}P_n / \mathrm{d}P ) \mathrm{d}P_n$.
Calculating the \ac{kl} divergence associated with the law of the Markov chain is impractical, since full knowledge of $P$ is required.
However, if convergence occurs, then the \emph{contrastive divergence} 
\begin{align}
    \Delta_n := D_{\mathrm{KL}}(P_n || P) - D_{\mathrm{KL}}(P_{n+1} || P) \label{eq: cd}
\end{align}
ought to vanish in the $n \rightarrow \infty$ limit.
Here we derive an explicit lower bound on the contrastive divergence, the maximisation of which we adopt as a heuristic to obtain a novel reward that is suitable for \ac{rlmh}.
In the setting of general Metropolis--Hastings algorithms, let $Q(\cdot | x)$ denote the proposal distribution with density $q(\cdot|x)$ assumed to exist for all $x \in \mathbb{R}^d$.
Let also $\alpha(x' | x)$ be the probability of accepting a proposed state $x'$ given the current state of the Markov chain is $x$.
The Markov transition kernel takes the form
\begin{equation} \label{eq: transition kernel}
    \mathsf{P}(A|x) = \int_A u(x'|x) \, \lambda(\mathrm{d}x') + r(x) \delta_{x}(A),
\end{equation}
where $u(x'|x) = \alpha(x'|x) q(x'|x)$, $r(x) = 1 - \int u(x'|x) \, \mathrm{d}\lambda(x')$, and $\delta_x$ is a point mass at $x \in \mathbb{R}^d$.
Since Metropolis--Hastings chains have a positive probability to reject, this transition kernel is not absolutely continuous with respect to Lebesgue measure $\lambda$.
To obtain a density for the transition kernel, we instead use the reference measure $\mu_x = \lambda + \delta_x$, for which
\begin{align}
\mathsf{p}(x'|x) := \frac{\mathrm{d}\mathsf{P}(\cdot|x)}{\mathrm{d}\mu_x}(x') = 
    \begin{cases} 
    u(x'|x), & \text{for } x' \neq x, \\
    r(x), & \text{for } x' = x.
    \end{cases}  \label{eq: phat}
\end{align}
The proof of the following result can be found in \Cref{app: proof thm 1}:

\begin{proposition}[Lower Bound for $\Delta_n$] 
\label{thm: cdlb}
Under the above assumptions, we obtain the following lower bound on the contrastive divergence:
\begin{equation} 
    \Delta_n \geq \E\left[\log\left(\frac{p(X_{n+1})}{p(X_{n}) \mathsf{p}(X_{n+1}|X_n)}\right) \right] 
    + \mathbb{E}\left[ \log\left( \frac{p_n(X_n)}{1 + p_n(X_{n+1})} \right) \right] .
    \label{eq: contrastive lower bound}
\end{equation}
\end{proposition}

\noindent The inequality in \Cref{thm: cdlb} is tight, since it is obtained from a single application of Jensen's inequality, which itself is tight.
Further, at stationarity the second term in \eqref{eq: contrastive lower bound} does not depend on the choice of Markov transition kernel $\mathsf{P}$. 

Our heuristic to deriving a novel reward for \ac{rlmh} is to maximise the relevant first term in the lower bound \eqref{eq: contrastive lower bound} with respect to the design of $\mathsf{P}$, which amounts to maximising the quantity
\begin{align}
\E\left[\log\left(\frac{p(X_{n+1})}{p(X_{n}) \mathsf{p}(X_{n+1}|X_n)}\right) \right] = \underbrace{ \E[\log p(X_{n+1}) - \log p(X_n)] }_{\text{exploitation}} + \underbrace{ \E[-\log \mathsf{p}(X_{n+1}|X_n)] }_{\text{exploration}} . \label{eq: ee}
\end{align}
The first term in \eqref{eq: ee} encourages exploitation, rewarding moves to regions of higher probability, while the second term in \eqref{eq: ee} encourages exploration, being the expected entropy of the transition kernel.
The exploration term admits the following more explicit lower bound, whose proof can be found in \Cref{app: proof second result}:

\begin{proposition}[Lower bound for exploration term in \eqref{eq: ee}] \label{prop: explicit}
Under the above assumptions we obtain a lower bound
    \begin{align}
        \hspace{-5pt} \underbrace{ \mathbb{E}[-\log \mathsf{p}(X_{n+1}|X_n) ] }_{\text{\normalfont exploration}} & \geq \mathbb{E}\left[ \begin{array}{l} - \left(1-  \alpha(X_{n+1}^* | X_n) \right) \log \left( 1 - \alpha(X_{n+1}^* | X_n)  \right)  \\
         \qquad - \alpha(X_{n+1}^*|X_n) [\log q(X_{n+1}^*|X_n) + \log \alpha(X_{n+1}^*|X_n)] \end{array}  \right] .   \label{eq: explore bound}
    \end{align}
\end{proposition}

\noindent Similarly to \Cref{thm: cdlb}, the bound in \Cref{prop: explicit} is also tight.
Let $\alpha_n := \alpha(X_{n+1}^\star | X_n)$ and $\bar{\alpha}_n = 1 - \alpha_n$.
Combining \eqref{eq: ee} and \eqref{eq: explore bound}, we obtain a reward that can be computed using only information available to the \ac{mdp} formulation of \ac{rlmh}
    \begin{align}
        r_n & := \underbrace{ \alpha_n [ \log p(X_{n+1}^\star) - \log p(X_n) ] }_{\text{exploitation}} + \underbrace{[  - \alpha_n \log \alpha_n - \bar{\alpha}_n \log\bar{\alpha}_n ]}_{\text{entropy of accept/reject}} + \underbrace{[ - \alpha_n \log q(X_{n+1}^\star | X_n ) ] }_{\approx \text{ ESJD}} \label{eq: our reward}
    \end{align}
which we call the \emph{contrastive divergence lower bound} (\acs{cdlb}).
The first term in \eqref{eq: our reward} is a Rao--Blackwellised unbiased estimator for the exploitation term in \eqref{eq: ee}, the second term is analogous to the \ac{aar} but can be computed in the context of an \ac{mdp}, and the final term is an affine transformation of the \ac{esjd} in the case of a Gaussian proposal.
This result may be of independent interest; informally, we have shown that \emph{joint} maximisation of (analogues of) the \ac{aar} and \ac{esjd}, together with an exploitation incentive, can be sufficient for optimising the performance of a Metropolis--Hastings kernel. 
In practice the \ac{cdlb} helps to avoid catastrophic failures that can sometimes occur in \ac{rlmh} during training with \ac{lesjd}; empirical support for this claim is provided in \Cref{sec: empirical}.

\section{Empirical Assessment}
\label{sec: empirical}

This section presents an empirical assessment of \ac{rlmh} applied to train \ac{rmala} based on the \ac{cdlb} reward.
Full details on the implementation of \ac{rmala} are contained in \Cref{ap: implement mala}, and full details for \ac{rlmh} are contained in \Cref{ap: implement}.
Code to reproduce our experiments is available via this \href{https://github.com/congyewang/Harnessing-the-Power-of-Reinforcement-Learning-for-Adaptive-MCMC}{GitHub link}.
All computation was performed on a Red Hat Enterprise Linux Server 7.9 (Maipo) system using Intel Xeon E5-2699 v4 CPUs and took 156 CPU hours in total.

\begin{figure}[t!]
\input{input/pic/demo_policy_2}
\caption{Training the Riemannian Metropolis-adjusted Langevin algorithm (RMALA) using \acf{rlmh}.
Top:  Target distributions $p$ to be sampled.
Middle:  Position-dependent step sizes $\epsilon(\cdot)$ learned using \ac{rlmh}.
Bottom:  Performance measured using \acf{mmd}.
A total of 10 independent simulations were performed and the 25, 50, and 75 percentiles of the MMD are displayed for position-independent \ac{rmala} (\textcolor{ggplot2}{blue}) as a function of the constant step size $\epsilon$, and for \ac{rlmh} (\textcolor{ggplot1}{red}) where $\epsilon(\cdot)$ is learned.
}
\label{fig: illustration}
\end{figure}

\paragraph{Illustration of Reinforcement Learning for \ac{rmala}}

First we illustrate the behaviour of \ac{rmala} trained using \ac{rlmh} on several examples in dimension $d = 2$, to better understand the nature of the policies that are being learned.
The top row of \Cref{fig: illustration} depicts the target distributions $p$ to be sampled.
Recall that we implement \ac{rmala} with preconditioner of the form $G(\cdot) = \epsilon(\cdot)^{-1} I$, with the step-size function $\epsilon(\cdot)$ learned using \ac{rlmh} based on the proposed \ac{cdlb} reward and, for these illustrations, $G_0$ was the identity matrix.
To visualise the learned policy, the middle row of \Cref{fig: illustration} plots the step size function $\epsilon(\cdot)$.
For the Laplace target in (a), our algorithm has learned to propose larger steps when the Markov chain is further from the mode; this is actually essential for the Markov chain to rapidly return from an excursion in the tail, since for this target $\nabla \log p$ is bounded.
In contrast, for the Gaussian target in (b) a near-constant step size $\epsilon(\cdot)$ has been learned, which we contend is optimal because for a Gaussian the Hessian of $\log p$ is constant.
For the banana target (c), the algorithm has learned to take smaller steps when in the narrow tails, which is sensible in light of the larger values of the gradient.
The bottom row of \Cref{fig: illustration} contrasts the performance of \ac{rlmh} to \ac{rmala} with a position-independent preconditioner $G(\cdot) = \epsilon^{-1} I$ across different values of the constant step size $\epsilon \in (0,\infty)$.
Performance was measured using \ac{mmd} based on the Gaussian kernel with unit lengthscale ($\ell = 1$); we precisely define \ac{mmd} in \Cref{app: mmd}.
Small improvements are observed when learning a position-dependent step size in (a) and (c), while this was not the case for (b) since here a constant step size is already optimal.

\paragraph{Framework for the Assessment}

\citet{magnusson2022posterior} introduced \texttt{posteriordb} to address the lack of a community benchmark for performance assessment in Bayesian computation.
By containing a range of realistic posterior distributions arising from genuine data analyses, together with gold-standard samples for many of these tasks, it aims to avoid the practices of hand-picked benchmarks, or classical problems that are insufficiently difficult.
Following \citet{wang2024reinforcement}, we considered the collection of 44 posteriors from \texttt{posteriordb} for which $10,000$ gold-standard samples are provided.
For each posterior and each sampling method, a total of $n = 30,000$ iterations were performed.
Assessment was based on the final $5,000$ iterations, during which no adaptation was permitted so that in the testing phase all Markov chains were $p$-invariant.
Performance was measured using \ac{mmd} relative to the gold-standard samples for each task.
Here the \ac{mmd} was based on the Gaussian kernel, whose lengthscale was determined using the \emph{median heuristic} applied to the gold standard samples \citep{garreau2017large}; c.f. \Cref{app: mmd}.

All algorithms that we consider require the preconditioner matrix $G_0$ to be estimated and this is not always straightforward; for example, one could attempt to approximate the Hessian matrix at a mode of the target, or use a simpler off-the-shelf adaptive \ac{mcmc} method in which a covariance matrix is iteratively learned \citep{andrieu2008tutorial,titsias2019gradient}.
Though it is possible to combine estimation of $G_0$  with estimation of $\epsilon$ or $\phi$, we sought to avoid confounding due to the specific choice of how $G_0$ is estimated; all algorithms that we compare take $G_0$ to be the inverse of the empirical covariance matrix of the $10^4$ gold-standard samples from the target.
This choice is justified since we report only a relative comparison of tuning strategies for \ac{rmala}.
Further, our results are robust to the use of alternative strategies for selecting $G_0$, as discussed in \Cref{app: alternative G0}.

\paragraph{Performance Assessment}

\begin{table}[t!]
    \centering
    \begin{adjustbox}{max width=\textwidth}
        \begin{tabular}{|cc|c|cc|cc|}
            \hline
            &    &  \textbf{Gradient-free RLMH}  & \multicolumn{2}{|c|}{\textbf{RMALA}} & \multicolumn{2}{|c|}{\textbf{RMALA-RLMH} (proposed)} \\
                Task  & $d$  &  LESJD  & AAR  & ESJD   & LESJD  & CDLB  \\
            \hline
            \rowcolor{gray!20}
            earnings-earn\_height & 3 & 1.8(0.1)E-1& 5.2(0.3)E-1 & 5.0(0.0)E-1 & \textbf{1.4(0.1)E-2} & 1.5(0.1)E-2 \\
            \rowcolor{gray!20}
            earnings-log10earn\_height & 3 & 1.6(0.0)E-1 & 1.4(0.0)E-2 & 1.5(0.1)E-2 & 1.4(0.1)E-2 & \textbf{1.3(0.0)E-2} \\
            \rowcolor{gray!20}
            earnings-logearn\_height & 3 & 1.6(0.0)E-1 & 1.5(0.0)E-2 & 1.6(0.1)E-2 & \textbf{1.3(0.1)E-2} & 1.4(0.0)E-2 \\
            gp\_pois\_regr-gp\_regr & 3 & 1.2(0.0)E-1 & \textbf{2.2(0.1)E-2} & 2.3(0.1)E-2 & 2.4(0.1)E-2 & 2.4(0.1)E-2 \\
            \rowcolor{gray!20}
            kidiq-kidscore\_momhs & 3 & 1.5(0.0)E-1 & 1.4(0.1)E-2 & 1.5(0.1)E-2 & \textbf{1.2(0.1)E-2} & \textbf{1.2(0.0)E-2} \\
            \rowcolor{gray!20}
            kidiq-kidscore\_momiq & 3 & 1.7(0.0)E-1 & 1.5(0.1)E-2 & 1.5(0.1)E-2 & \textbf{1.2(0.0)E-2} & 1.4(0.1)E-2 \\
            \rowcolor{gray!20}
            kilpisjarvi\_mod-kilpisjarvi & 3 & 1.7(0.0)E-1 & 5.9(0.3)E-1 & 6.0(0.3)E-1 & \textbf{2.5(0.7)E-2} & \textbf{2.5(0.6)E-2} \\
            \rowcolor{gray!20}
            mesquite-logmesquite\_logvolume & 3 & 1.3(0.0)E-1 & 2.2(0.1)E-2 & 2.4(0.1)E-2 & \textbf{1.9(0.1)E-2} & 2.0(0.1)E-2 \\
            \rowcolor{gray!20}
            arma-arma11 & 4 &  1.2(0.0)E-1 & 2.3(0.1)E-2 & 2.5(0.1)E-2 & 2.1(0.1)E-2 & \textbf{2.0(0.1)E-2} \\
            \rowcolor{gray!20}
            earnings-logearn\_height\_male & 4 & 1.6(0.0)E-1 & 1.4(0.1)E-2 & 1.6(0.1)E-2 & \textbf{1.3(0.1)E-2} & \textbf{1.3(0.1)E-2} \\
            \rowcolor{gray!20}
            earnings-logearn\_logheight\_male & 4 & 1.6(0.0)E-1 & 1.3(0.1)E-2 & 1.5(0.1)E-2 & \textbf{1.1(0.1)E-2} & 1.2(0.1)E-2 \\
            garch-garch11 & 4 &  1.4(0.0)E-1 & \textbf{1.4(0.1)E-2} & 1.7(0.1)E-2 &  \cellcolor{black!50} - & 8.9(7.2)E-2 \\
            \rowcolor{gray!20}
            hmm\_example-hmm\_example & 4 &  1.3(0.0)E-1 & 2.1(0.1)E-2 & 2.2(0.1)E-2 & \textbf{1.7(0.1)E-2} & 1.9(0.1)E-2 \\
            \rowcolor{gray!20}
            kidiq-kidscore\_momhsiq & 4 &  1.4(0.0)E-1 & 1.6(0.1)E-2 & 1.8(0.1)E-2 & \textbf{1.2(0.1)E-2} & 1.7(0.1)E-2 \\
            \vdots & \vdots & \vdots & \vdots & \vdots & \vdots & \vdots \\
            diamonds-diamonds & 26 &  2.0(0.1)E0 & 6.1(0.8)E-2 & \textbf{5.3(0.6)E-2} & 6.2(0.6)E-2 & 5.8(0.7)E-2 \\
            mcycle\_gp-accel\_gp & 66 &  1.9(0.0)E0 & \textbf{7.6(0.0)E-1} & 7.7(0.0)E-1 & \cellcolor{black!50} - & 7.7(0.0)E-1 \\
            \hline
        \end{tabular}
    \end{adjustbox}

\medskip
    
    \caption{Benchmarking using \texttt{posteriordb}. 
  Here we compared standard \ac{rmala} with constant step-size $\epsilon \in (0,\infty)$ tuned either by matching the \acf{aar} to 0.574 or by maximising the \acf{esjd}, to \ac{rmala} with position-dependent step size $\epsilon(\cdot)$ learned using \ac{rlmh}, based either using the \ac{lesjd} reward of \citet{wang2024reinforcement} or the proposed \ac{cdlb}. 
  The gradient-free \ac{rlmh} method of \citet{wang2024reinforcement} is also included as a benchmark.
  Performance was measured using \acf{mmd} based on the Gaussian kernel with lengthscale selected using the median heuristic applied to $10,000$ gold-standard samples from the target, and $d$ is the dimension of the target. 
  Results are based on an average of 10 replicates, with standard errors (in parentheses) reported.
  The method with smallest average \ac{mmd} is highlighted in \textbf{bold}, gray rows indicate tasks for which \ac{rlmh} performed best, and dashes indicate where catastrophic failure occurred during training of \ac{rlmh}.
  Full results are in \Cref{app: pdb full}.}
  \label{tab: posteriordb shortened}
\end{table}

Using the \texttt{posteriordb} benchmark, we compared the performance of standard \ac{rmala}, i.e. based on a preconditioner of the form $G(\cdot) = \epsilon^{-1} G_0$, with constant step-size $\epsilon$ tuned either by matching the \acf{aar} to 0.574 or by maximising the \acf{esjd}, to \ac{rmala} with preconditioner of the form $G(\cdot) = \epsilon(\cdot)^{-1} G_0$ with $\epsilon(\cdot)$ learned using \ac{rlmh}, based either using the \ac{lesjd} reward of \citet{wang2024reinforcement} or the proposed \acf{cdlb}. 
The gradient-free \ac{rlmh} method of \citet{wang2024reinforcement} was also included as a benchmark.
Full implementation details for \ac{rmala} are contained in \Cref{ap: implement mala}, while full implementation details for \ac{rlmh} are contained in \Cref{ap: implement}.
The computational cost of all gradient-based algorithms was equal, in terms of the number of calls to evaluate either the target $p$ or its gradient.

\begin{wrapfigure}{r}{0.4\linewidth}
    \centering
    \input{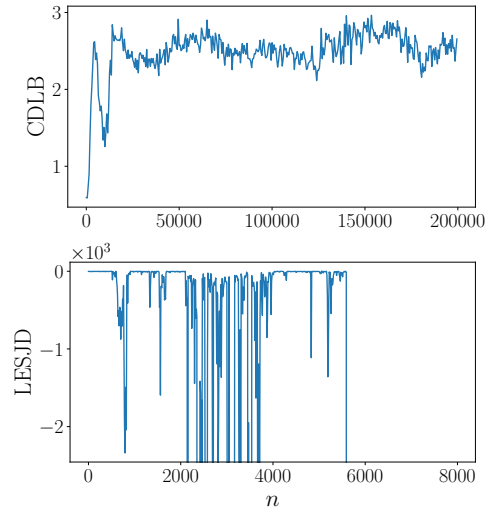}
    \caption{Comparison of training based on \ac{cdlb} (top panel) and \ac{lesjd} (bottom panel).  
    Moving averages of the rewards $r_n$ are displayed.
    Training with \ac{cdlb} oscillates during the initial exploration phase, before stabilising once sufficient information has been gained. 
    In contrast, \ac{lesjd} provides insufficient signal and can lead to catastrophic failure of \ac{rlmh}.
    }
    \label{fig:reward plot}
\end{wrapfigure}

Summarised results are shown in \Cref{tab: posteriordb shortened}, with full results presented in \Cref{app: pdb full}.
The headline is that a position-dependent step size $\epsilon(\cdot)$, learned using \ac{rlmh}, outperformed a (optimised) constant step size in \ac{rmala}, with the smallest average \ac{mmd} achieved by \ac{rlmh} in 89\% of the tasks considered.
All gradient-based samplers outperformed the gradient-free instance of \ac{rlmh} studied in \citet{wang2024reinforcement}.
On the negative side, for the highest-dimensional tasks no improvement due to the position-dependent step size was observed; this can be attributed to increased difficulty of learning the step size function $\epsilon(\cdot)$ when the domain is high-dimensional.
Further, a closer inspection of the sample paths reveals that algorithms based on \ac{rlmh} converge in a more complex manner compared to classical adaptive \ac{mcmc}, for which performance typically improves monotonically when a constant step size $\epsilon$ is being optimised.
This behaviour is illustrated in the top panel of \Cref{fig:reward plot}.

A secondary, but still important result was that, although the \ac{lesjd} reward tended to perform slightly better than \ac{cdlb} as a training objective for \ac{rlmh}, catastrophic failures were encountered during training for 7\% of the tasks when the \ac{lesjd} reward was used.
Such an instance of catastrophic failure is documented in the bottom panel of \Cref{fig:reward plot}.
These failures were completely ameliorated when the \ac{cdlb} reward was used.

Though we focused on \ac{rmala}, the prototypical gradient-based sampling method, it is interesting to explore the potential of \ac{rlmh} for tuning of other gradient-based samplers.
To this end, analogous results for the Barker proposal of \citet{livingstone2022barker} can be found in \Cref{ap: Barker}.
Interestingly we did not find compelling evidence for the use of a flexible step size for the Barker proposal; this makes sense because the Barker proposal was specifically developed to trade-off some efficiency for being relatively insensitive to the step size of the proposal.

\section{Discussion}
\label{sec: discuss}

The impressive performance of modern \ac{rl}, on applications such as autonomous driving \citep{kiran2021deep}, gaming \citep{silver2016mastering}, and natural language processing \citep{shinn2024reflexion}, serves as strong motivation for investigating if and how techniques from \ac{rl} can be harnessed for adaptive \ac{mcmc}.
This paper is the first to explore the use of \ac{rl} to tune gradient-based \ac{mcmc}.
Our empirical findings demonstrate that fast-mixing Markov kernels can be effectively learned within the framework of \ac{rlmh}.
The main limitation is that poor sample quality can occur during the initial exploration phase of \ac{rl}, analogous to an extended warm-up period in \ac{mcmc}; we therefore recommend the use of gradient-based \ac{rlmh} primarily in settings where high-quality posterior approximation is required (i.e. for fast posterior approximations at lower quality, simpler methods should be preferred).
In addition we presented the novel \ac{cdlb} reward, which was shown to stabilise training in \ac{rlmh} and may have applications in adaptive \ac{mcmc} beyond this work.

For future research, we highlight that learning the proposal covariance structure (i.e. $G_0$ in the setting of \ac{rmala}) is an open challenge for \ac{rl}; we speculate that reduced-rank covariance matrix approximations may be useful here, enabling the difficulty of the learning task to be reduced, but our attempts to implement this (not shown) were unsuccessful.
Since multiple instances of \ac{mcmc} can be run in parallel, federated Q-learning and related strategies may offer a promising way forward.
Alternatively, one could explore the application of \ac{rl} to tuning other sampling methods, such as selecting the step size and number of integrator `leaps' in adaptive Hamiltonian Monte Carlo \citep{wang2013adaptive,hoffman2014no,christiansen2023self}, or consider the formulation of related algorithms such as multiple-try Metropolis \citep{liu2000multiple,doig2025unified} and delayed acceptance \ac{mcmc} \citep{christen2005markov} as \acp{mdp} amenable to \ac{rl}.

\paragraph{Acknowledgements} 
CW was supported by the China Scholarship Council under Grant Number 202208890004. 
HK and CJO were supported by the Engineering and Physical Sciences Research Council EP/W019590/1.
CJO was supported by a Philip Leverhulme Prize PLP-2023-004.

\bibliographystyle{abbrvnat}
\bibliography{bibliography}

\begin{thebibliography}{55}
\providecommand{\natexlab}[1]{#1}
\providecommand{\url}[1]{\texttt{#1}}
\expandafter\ifx\csname urlstyle\endcsname\relax
  \providecommand{\doi}[1]{doi: #1}\else
  \providecommand{\doi}{doi: \begingroup \urlstyle{rm}\Url}\fi

\bibitem[Andrieu and Moulines(2006)]{andrieu2006ergodicity}
C.~Andrieu and {\'E}.~Moulines.
\newblock On the ergodicity properties of some adaptive {MCMC} algorithms.
\newblock \emph{The Annals of Applied Probability}, 16\penalty0 (1):\penalty0 1462--1505, 2006.

\bibitem[Andrieu and Robert(2001)]{andrieu2001controlled}
C.~Andrieu and C.~P. Robert.
\newblock Controlled {MCMC} for optimal sampling.
\newblock Technical Report~33, Center for Research in Economics and Statistics, 2001.

\bibitem[Andrieu and Thoms(2008)]{andrieu2008tutorial}
C.~Andrieu and J.~Thoms.
\newblock A tutorial on adaptive {MCMC}.
\newblock \emph{Statistics and Computing}, 18:\penalty0 343--373, 2008.

\bibitem[Arbel et~al.(2021)Arbel, Matthews, and Doucet]{arbel2021annealed}
M.~Arbel, A.~Matthews, and A.~Doucet.
\newblock Annealed flow transport {M}onte {C}arlo.
\newblock In \emph{Proceedings of the 38th International Conference on Machine Learning}, 2021.

\bibitem[Atchade et~al.(2011)Atchade, Fort, Moulines, and Priouret]{atchade2011adaptive}
Y.~Atchade, G.~Fort, E.~Moulines, and P.~Priouret.
\newblock \emph{Adaptive {M}arkov chain {M}onte {C}arlo: {T}heory and methods}.
\newblock Bayesian Time Series Models. Cambridge University Press, 2011.

\bibitem[Atchad{\'e} and Rosenthal(2005)]{atchade2005adaptive}
Y.~F. Atchad{\'e} and J.~S. Rosenthal.
\newblock On adaptive {M}arkov chain {M}onte {C}arlo algorithms.
\newblock \emph{Bernoulli}, 11\penalty0 (5):\penalty0 815--828, 2005.

\bibitem[Barker(1965)]{barker1965monte}
A.~A. Barker.
\newblock Monte {C}arlo calculations of the radial distribution functions for a proton-electron plasma.
\newblock \emph{Australian Journal of Physics}, 18\penalty0 (2):\penalty0 119--134, 1965.

\bibitem[Biron-Lattes et~al.(2024)Biron-Lattes, Surjanovic, Syed, Campbell, and Bouchard-C{\^o}t{\'e}]{biron2023automala}
M.~Biron-Lattes, N.~Surjanovic, S.~Syed, T.~Campbell, and A.~Bouchard-C{\^o}t{\'e}.
\newblock auto{MALA}: {L}ocally adaptive {M}etropolis-adjusted {L}angevin algorithm.
\newblock In \emph{Proceedings of the 27th International Conference on Artificial Intelligence and Statistics}, 2024.

\bibitem[Blessing et~al.(2025)Blessing, Berner, Richter, and Neumann]{blessing2025underdamped}
D.~Blessing, J.~Berner, L.~Richter, and G.~Neumann.
\newblock Underdamped diffusion bridges with applications to sampling.
\newblock In \emph{Proceedings of the 13th International Conference on Learning Representations}, 2025.

\bibitem[Bui-Thanh and Ghattas(2012)]{bui2012scaled}
T.~Bui-Thanh and O.~Ghattas.
\newblock A scaled stochastic {N}ewton algorithm for {M}arkov chain {M}onte {C}arlo simulations.
\newblock \emph{SIAM Journal on Uncertainty Quantification}, pages 1--25, 2012.

\bibitem[Chen et~al.(2023)Chen, Huang, Huang, Reich, and Stuart]{chen2023sampling}
Y.~Chen, D.~Z. Huang, J.~Huang, S.~Reich, and A.~M. Stuart.
\newblock Sampling via gradient flows in the space of probability measures.
\newblock \emph{arXiv preprint arXiv:2310.03597}, 2023.

\bibitem[Chopin and Papaspiliopoulos(2020)]{chopin2020introduction}
N.~Chopin and O.~Papaspiliopoulos.
\newblock \emph{An Introduction to Sequential Monte Carlo}.
\newblock Springer, 2020.

\bibitem[Christen and Fox(2005)]{christen2005markov}
J.~A. Christen and C.~Fox.
\newblock Markov chain {M}onte {C}arlo using an approximation.
\newblock \emph{Journal of Computational and Graphical statistics}, 14\penalty0 (4):\penalty0 795--810, 2005.

\bibitem[Christiansen et~al.(2023)Christiansen, Errica, and Alesiani]{christiansen2023self}
H.~Christiansen, F.~Errica, and F.~Alesiani.
\newblock Self-tuning {H}amiltonian {M}onte {C}arlo for accelerated sampling.
\newblock \emph{The Journal of Chemical Physics}, 159\penalty0 (23), 2023.

\bibitem[Chung et~al.(2023)Chung, Leung, Pun, and Zhang]{chung2020multi}
E.~Chung, W.~T. Leung, S.-M. Pun, and Z.~Zhang.
\newblock Multi-agent reinforcement learning accelerated {MCMC} on multiscale inversion problem.
\newblock \emph{Journal of Scientific Computing}, 93\penalty0 (2), 2023.

\bibitem[Coullon et~al.(2023)Coullon, South, and Nemeth]{coullon2023efficient}
J.~Coullon, L.~South, and C.~Nemeth.
\newblock Efficient and generalizable tuning strategies for stochastic gradient {MCMC}.
\newblock \emph{Statistics and Computing}, 33\penalty0 (3):\penalty0 66, 2023.

\bibitem[Dharamshi et~al.(2023)Dharamshi, Ngo, and Rosenthal]{dharamshi2023sampling}
A.~Dharamshi, V.~Ngo, and J.~S. Rosenthal.
\newblock Sampling by divergence minimization.
\newblock \emph{Communications in Statistics}, pages 1--25, 2023.

\bibitem[Doig and Wang(2025)]{doig2025unified}
R.~Doig and L.~Wang.
\newblock A unified framework for multiple-try {M}etropolis algorithms.
\newblock \emph{arXiv preprint arXiv:2503.11583}, 2025.

\bibitem[Eberle(2016)]{eberle2016reflection}
A.~Eberle.
\newblock Reflection couplings and contraction rates for diffusions.
\newblock \emph{Probability Theory and Related Fields}, 166:\penalty0 851--886, 2016.

\bibitem[Fox and Roberts(2012)]{fox2012tutorial}
C.~W. Fox and S.~J. Roberts.
\newblock A tutorial on variational {B}ayesian inference.
\newblock \emph{Artificial Intelligence Review}, 38:\penalty0 85--95, 2012.

\bibitem[Fujimoto et~al.(2018)Fujimoto, Hoof, and Meger]{fujimoto2018addressing}
S.~Fujimoto, H.~Hoof, and D.~Meger.
\newblock Addressing function approximation error in actor-critic methods.
\newblock In \emph{Proceedings of the 37th International Conference on Machine Learning}, 2018.

\bibitem[Garreau et~al.(2017)Garreau, Jitkrittum, and Kanagawa]{garreau2017large}
D.~Garreau, W.~Jitkrittum, and M.~Kanagawa.
\newblock Large sample analysis of the median heuristic.
\newblock \emph{arXiv preprint arXiv:1707.07269}, 2017.

\bibitem[Gelman et~al.(1997)Gelman, Gilks, and Roberts]{gelman1997weak}
A.~Gelman, W.~R. Gilks, and G.~O. Roberts.
\newblock Weak convergence and optimal scaling of random walk {M}etropolis algorithms.
\newblock \emph{The Annals of Applied Probability}, 7\penalty0 (1):\penalty0 110--120, 1997.

\bibitem[Girolami and Calderhead(2011)]{girolami2011riemann}
M.~Girolami and B.~Calderhead.
\newblock Riemann manifold {L}angevin and {H}amiltonian {M}onte {C}arlo methods.
\newblock \emph{Journal of the Royal Statistical Society Series B: Statistical Methodology}, 73\penalty0 (2):\penalty0 123--214, 2011.

\bibitem[Haario et~al.(2006)Haario, Laine, Mira, and Saksman]{haario2006dram}
H.~Haario, M.~Laine, A.~Mira, and E.~Saksman.
\newblock {DRAM}: {E}fficient adaptive {MCMC}.
\newblock \emph{Statistics and Computing}, 16:\penalty0 339--354, 2006.

\bibitem[Hoffman and Gelman(2014)]{hoffman2014no}
M.~D. Hoffman and A.~Gelman.
\newblock The {No-U-Turn} {S}ampler: {A}daptively setting path lengths in {H}amiltonian {M}onte {C}arlo.
\newblock \emph{Journal of Machine Learning Research}, 15\penalty0 (1):\penalty0 1593--1623, 2014.

\bibitem[Hunt-Smith et~al.(2024)Hunt-Smith, Melnitchouk, Ringer, Sato, Thomas, and White]{hunt2024accelerating}
N.~T. Hunt-Smith, W.~Melnitchouk, F.~Ringer, N.~Sato, A.~W. Thomas, and M.~J. White.
\newblock Accelerating {M}arkov chain {M}onte {C}arlo sampling with diffusion models.
\newblock \emph{Computer Physics Communications}, 296:\penalty0 109059, 2024.

\bibitem[Kaelbling et~al.(1996)Kaelbling, Littman, and Moore]{kaelbling1996reinforcement}
L.~P. Kaelbling, M.~L. Littman, and A.~W. Moore.
\newblock Reinforcement learning: {A} survey.
\newblock \emph{Journal of Artificial Intelligence Research}, 4:\penalty0 237--285, 1996.

\bibitem[Kim et~al.(2025)Kim, Xu, Gardner, and Campbell]{kim2025tuning}
K.~Kim, Z.~Xu, J.~R. Gardner, and T.~Campbell.
\newblock Tuning sequential {M}onte {C}arlo samplers via greedy incremental divergence minimization.
\newblock \emph{arXiv preprint arXiv:2503.15704}, 2025.

\bibitem[Kiran et~al.(2021)Kiran, Sobh, Talpaert, Mannion, Al~Sallab, Yogamani, and P{\'e}rez]{kiran2021deep}
B.~R. Kiran, I.~Sobh, V.~Talpaert, P.~Mannion, A.~A. Al~Sallab, S.~Yogamani, and P.~P{\'e}rez.
\newblock Deep reinforcement learning for autonomous driving: A survey.
\newblock \emph{IEEE Transactions on Intelligent Transportation Systems}, 23\penalty0 (6):\penalty0 4909--4926, 2021.

\bibitem[Lillicrap et~al.(2016)Lillicrap, Hunt, Pritzel, Heess, Erez, Tassa, Silver, and Wierstra]{lillicrap2015continuous}
T.~P. Lillicrap, J.~J. Hunt, A.~Pritzel, N.~Heess, T.~Erez, Y.~Tassa, D.~Silver, and D.~Wierstra.
\newblock Continuous control with deep reinforcement learning.
\newblock In \emph{Proceedings of the 4th International Conference on Learning Representations}, 2016.

\bibitem[Liu et~al.(2000)Liu, Liang, and Wong]{liu2000multiple}
J.~S. Liu, F.~Liang, and W.~H. Wong.
\newblock The multiple-try method and local optimization in {M}etropolis sampling.
\newblock \emph{Journal of the American Statistical Association}, 95\penalty0 (449):\penalty0 121--134, 2000.

\bibitem[Liu et~al.(2024)Liu, Surjanovic, Biron-Lattes, Bouchard-C{\^o}t{\'e}, and Campbell]{liu2024autostep}
T.~Liu, N.~Surjanovic, M.~Biron-Lattes, A.~Bouchard-C{\^o}t{\'e}, and T.~Campbell.
\newblock {AutoStep}: {L}ocally adaptive involutive {MCMC}.
\newblock \emph{arXiv preprint arXiv:2410.18929}, 2024.

\bibitem[Livingstone and Zanella(2022)]{livingstone2022barker}
S.~Livingstone and G.~Zanella.
\newblock The {B}arker proposal: {C}ombining robustness and efficiency in gradient-based {MCMC}.
\newblock \emph{Journal of the Royal Statistical Society Series B: Statistical Methodology}, 84\penalty0 (2):\penalty0 496--523, 2022.

\bibitem[Magnusson et~al.(2025)Magnusson, Torgander, B\"{u}rkner, Zhang, Carpenter, and Vehtari]{magnusson2022posterior}
M.~Magnusson, J.~Torgander, P.-C. B\"{u}rkner, L.~Zhang, B.~Carpenter, and A.~Vehtari.
\newblock \texttt{posteriordb}: Testing, benchmarking and developing bayesian inference algorithms.
\newblock In \emph{Proceedings of the 28th International Conference on Artificial Intelligence and Statistics}, 2025.
\newblock URL \url{https://github.com/stan-dev/posteriordb}.

\bibitem[Mahendran et~al.(2012)Mahendran, Wang, Hamze, and De~Freitas]{mahendran2012adaptive}
N.~Mahendran, Z.~Wang, F.~Hamze, and N.~De~Freitas.
\newblock Adaptive {MCMC} with {B}ayesian optimization.
\newblock In \emph{Proceedings of the 15th International Conference on Artificial Intelligence and Statistics}, pages 751--760, 2012.

\bibitem[Naik et~al.(2024)Naik, Wan, Tomar, and Sutton]{naik2024rewardcentering}
A.~Naik, Y.~Wan, M.~Tomar, and R.~S. Sutton.
\newblock Reward centering, 2024.
\newblock URL \url{https://arxiv.org/abs/2405.09999}.

\bibitem[Pasarica and Gelman(2010)]{pasarica2010adaptively}
C.~Pasarica and A.~Gelman.
\newblock Adaptively scaling the {M}etropolis algorithm using expected squared jumped distance.
\newblock \emph{Statistica Sinica}, pages 343--364, 2010.

\bibitem[Potter and Swendsen(2015)]{potter20150}
C.~C. Potter and R.~H. Swendsen.
\newblock 0.234: {T}he myth of a universal acceptance ratio for {M}onte {C}arlo simulations.
\newblock \emph{Physics Procedia}, 68:\penalty0 120--124, 2015.

\bibitem[Rezende and Mohamed(2015)]{rezende2015variational}
D.~Rezende and S.~Mohamed.
\newblock Variational inference with normalizing flows.
\newblock In \emph{Proceedings of the 32nd International Conference on Machine Learning}, 2015.

\bibitem[Robbins and Monro(1951)]{robbins1951stochastic}
H.~Robbins and S.~Monro.
\newblock A stochastic approximation method.
\newblock \emph{The Annals of Mathematical Statistics}, 22\penalty0 (3):\penalty0 400--407, 1951.
\newblock \doi{10.1214/aoms/1177729586}.

\bibitem[Roberts and Rosenthal(1998)]{roberts1998optimal}
G.~O. Roberts and J.~S. Rosenthal.
\newblock Optimal scaling of discrete approximations to {L}angevin diffusions.
\newblock \emph{Journal of the Royal Statistical Society Series B: Statistical Methodology}, 60\penalty0 (1):\penalty0 255--268, 1998.

\bibitem[Roberts and Rosenthal(2007)]{roberts2007coupling}
G.~O. Roberts and J.~S. Rosenthal.
\newblock Coupling and ergodicity of adaptive {M}arkov chain {M}onte {C}arlo algorithms.
\newblock \emph{Journal of Applied Probability}, 44\penalty0 (2):\penalty0 458--475, 2007.

\bibitem[Roberts and Rosenthal(2009)]{roberts2009examples}
G.~O. Roberts and J.~S. Rosenthal.
\newblock Examples of adaptive {MCMC}.
\newblock \emph{Journal of Computational and Graphical Statistics}, 18\penalty0 (2):\penalty0 349--367, 2009.

\bibitem[Roberts and Stramer(2002)]{roberts2002langevin}
G.~O. Roberts and O.~Stramer.
\newblock Langevin diffusions and metropolis-hastings algorithms.
\newblock \emph{Methodology and Computing in Applied Probability}, 4:\penalty0 337--357, 2002.

\bibitem[Roberts and Tweedie(1996)]{robert1996exponential}
G.~O. Roberts and R.~Tweedie.
\newblock Exponential convergence of {L}angevin diffusions and their discrete approximation.
\newblock \emph{Bernoulli}, 2:\penalty0 341--363, 1996.

\bibitem[Roy and Zhang(2023)]{roy2023convergence}
V.~Roy and L.~Zhang.
\newblock Convergence of position-dependent {MALA} with application to conditional simulation in {GLMMs}.
\newblock \emph{Journal of Computational and Graphical Statistics}, 32\penalty0 (2):\penalty0 501--512, 2023.

\bibitem[Sherlock and Roberts(2009)]{sherlock2009optimal}
C.~Sherlock and G.~O. Roberts.
\newblock Optimal scaling of the random walk {M}etropolis on elliptically symmetric unimodal targets.
\newblock \emph{Bernoulli}, 15\penalty0 (3):\penalty0 774--798, 2009.

\bibitem[Shinn et~al.(2023)Shinn, Cassano, Gopinath, Narasimhan, and Yao]{shinn2024reflexion}
N.~Shinn, F.~Cassano, A.~Gopinath, K.~Narasimhan, and S.~Yao.
\newblock Reflexion: Language agents with verbal reinforcement learning.
\newblock In \emph{Proceedings of the 36th Conference on Neural Information Processing Systems}, 2023.

\bibitem[Silver et~al.(2014)Silver, Lever, Heess, Degris, Wierstra, and Riedmiller]{silver2014deterministic}
D.~Silver, G.~Lever, N.~Heess, T.~Degris, D.~Wierstra, and M.~Riedmiller.
\newblock Deterministic policy gradient algorithms.
\newblock In \emph{Proceedings of the 31st International Conference on Machine Learning}, pages 387--395. PMLR, 2014.

\bibitem[Silver et~al.(2016)Silver, Huang, Maddison, Guez, Sifre, Van Den~Driessche, Schrittwieser, Antonoglou, Panneershelvam, Lanctot, Dieleman, Grewe, Nham, Kalchbrenner, Sutskever, Lillicrap, Leach, Kavukcuoglu, Graepel, and Hassabis]{silver2016mastering}
D.~Silver, A.~Huang, C.~J. Maddison, A.~Guez, L.~Sifre, G.~Van Den~Driessche, J.~Schrittwieser, I.~Antonoglou, V.~Panneershelvam, M.~Lanctot, S.~Dieleman, D.~Grewe, J.~Nham, N.~Kalchbrenner, I.~Sutskever, T.~Lillicrap, M.~Leach, K.~Kavukcuoglu, T.~Graepel, and D.~Hassabis.
\newblock Mastering the game of {G}o with deep neural networks and tree search.
\newblock \emph{Nature}, 529\penalty0 (7587):\penalty0 484--489, 2016.

\bibitem[Titsias and Dellaportas(2019)]{titsias2019gradient}
M.~Titsias and P.~Dellaportas.
\newblock Gradient-based adaptive {M}arkov chain {M}onte {C}arlo.
\newblock In \emph{Proceedings of the 32nd Conference on Neural Information Processing Systems}, 2019.

\bibitem[Wang et~al.(2025)Wang, Chen, Kanagawa, Oates, et~al.]{wang2024reinforcement}
C.~Wang, W.~Chen, H.~Kanagawa, C.~Oates, et~al.
\newblock Reinforcement learning for adaptive {MCMC}.
\newblock In \emph{Proceedings of the 28th International Conference on Artificial Intelligence and Statistics}, 2025.

\bibitem[Wang et~al.(2013)Wang, Mohamed, and Freitas]{wang2013adaptive}
Z.~Wang, S.~Mohamed, and N.~Freitas.
\newblock Adaptive {H}amiltonian and {R}iemann manifold {M}onte {C}arlo.
\newblock In \emph{Proceedings of the 30th International Conference on Machine Learning}, 2013.

\bibitem[Wang et~al.(2021)Wang, Xia, Lyu, and Ling]{wang2021reinforcement}
Z.~Wang, Y.~Xia, S.~Lyu, and C.~Ling.
\newblock Reinforcement learning-aided {M}arkov chain {M}onte {C}arlo for lattice {G}aussian sampling.
\newblock In \emph{Proceedings of the IEEE Information Theory Workshop}. IEEE, 2021.

\end{thebibliography}

\appendix

These appendices supplement the paper \emph{Harnessing the Power of Reinforcement Learning for Adaptive MCMC}.
\Cref{app: proof thm 1} contains the Proof of \Cref{thm: cdlb}.
\Cref{app: proof second result} contains the proof of \Cref{prop: explicit}.
Full implementational details for \ac{rmala} and \ac{rlmh} are contained, respectively, in \Cref{ap: implement mala} and \Cref{ap: implement}.
Full results from the \texttt{posteriordb} benchmark are contained in \Cref{subsec: full results}.

\section{Proof of \Cref{thm: cdlb}}
\label{app: proof thm 1}

This appendix contains the proof of \Cref{thm: cdlb} in the main text.

\begin{proof}[Proof of \Cref{thm: cdlb}]
    First, to clarify the notation from \eqref{eq: phat}, note that the (Lebesgue) density $p_{n+1}$ for the law $P_{n+1}$ of the $(n+1)$th state $X_{n+1}$ of the Markov chain can be expressed recursively as
    \begin{align}    
        p_{n+1}(x) = \underbrace{ r(x) p_n(x) }_{\text{reject}} + \underbrace{ \int u(x | x') p_n(x') \, \mathrm{d}\lambda(x') }_{\text{accept}} 
        = \int \mathsf{p}(x|x') p_n(x') \; \mathrm{d}\mu_{x}(x'). \label{eq: identity}
    \end{align}
    Next, we manipulate the definition of contrastive divergence $\Delta_n$ in, stated in \eqref{eq: cd}, to obtain that
    \begin{align}
        \Delta_n  & = \int \log\left( \frac{\mathrm{d}P_n}{\mathrm{d}P} \right) \mathrm{d}P_n - \int \log\left( \frac{\mathrm{d}P_{n+1}}{\mathrm{d}P} \right) \mathrm{d}P_{n+1} \nonumber \\
        &= \int p_n \log p_n  \,\mathrm{d}\lambda - \int p_n \log p \,\mathrm{d}\lambda + \int p_{n+1} \log p \,\mathrm{d}\lambda - \int p_{n+1} \log p_{n+1} \, \mathrm{d}\lambda  \nonumber \\
        & = D_{\mathrm{KL}}(P_n || \lambda) + \mathbb{E}\left[ \log\left( \frac{p(X_{n+1})}{p(X_n)} \right) \right] - \int p_{n+1} \log p_{n+1} \, \mathrm{d}\lambda . \label{eq: interm der}
    \end{align}
    Considering the final integral, we use \eqref{eq: identity} to obtain
    \begin{align*}
        \int p_{n+1} \log p_{n+1} \, \mathrm{d}\lambda
        &= \int \left(\int \mathsf{p}(x|x') p_n(x')\, \mathrm{d}\mu_{x}(x')\right) \log\left(\int \mathsf{p}(x|x') p_n(x')\, \textbf{}\mu_{x}(x')\right)\, \mathrm{d}\lambda(x) \\
        & = \int \underbrace{ \psi\left( \int \mathsf{p}(x|x') p_n(x')\, \mathrm{d}\mu_{x}(x') \right) }_{(\star)} \, \mathrm{d}\lambda(x)
    \end{align*}
    where $\psi(y) := y \log y$.
    Next we will use Jensen's inequality on $(\star)$, and to make this clear we introduce an explicit normalisation constant
    $$
    Z_n(x) := \int p_n(x') \dd \mu_x(x') = 1 + p_n(x),
    $$
    so
    \begin{align*}
        \int \mathsf{p}(x|x') p_n(x')\, \mathrm{d}\mu_{x}(x') = \int [ Z_n(x) \mathsf{p}(x|x') ] \frac{p_n(x')}{Z_n(x)}\, \mathrm{d}\mu_{x}(x'), \qquad \int \frac{p_n(x')}{Z_n(x)}\, \mathrm{d}\mu_{x}(x') = 1 .
    \end{align*}
    Since $\psi$ is convex, Jensen's inequality gives that
    \begin{align*}
        (\star) = \psi\left( \int [ Z_n(x) \mathsf{p}(x|x') ] \frac{p_n(x')}{Z_n(x)}\, \mathrm{d}\mu_{x}(x') \right) 
        & \leq \int \psi( Z_n(x) \mathsf{p}(x|x')  ) \frac{p_n(x')}{Z_n(x)} \, \mathrm{d}\mu_x(x')  \\
        & = \int \mathsf{p}(x | x') p_n(x') \log[ Z_n(x) \mathsf{p}(x|x') ]  \, \mathrm{d}\mu_x(x') 
    \end{align*}
    from which we deduce that
    \begin{align}
        \int p_{n+1} \log p_{n+1} \, \mathrm{d}\lambda 
        & \leq \mathbb{E}[\log \mathsf{p}(X_{n+1}|X_n)] + \mathbb{E}[\log Z_n(X_{n+1})]. \label{eq: ineq}
    \end{align}    
    Substituting \eqref{eq: ineq} into \eqref{eq: interm der} gives that
    \begin{align*}
        \Delta_n & \geq D_{\mathrm{KL}}(P_n || \lambda) + \mathbb{E}\left[ \log\left( \frac{p(X_{n+1})}{p(X_n)} \right) \right] - \mathbb{E}[\log \mathsf{p}(X_{n+1}|X_n)] - \mathbb{E}[\log Z_n(X_{n+1})] \\
        & = \E\left[\log\left(\frac{p(X_{n+1})}{p(X_{n}) \mathsf{p}(X_{n+1}|X_n)}\right) \right] 
          + \mathbb{E}\left[ \log\left( \frac{p_n(X_n)}{1 + p_n(X_{n+1})} \right) \right]
    \end{align*}
    and completes the argument.
\end{proof}

\section{Proof of \Cref{prop: explicit}}
\label{app: proof second result}

This appendix contains the proof of \Cref{prop: explicit} in the main text.

\begin{proof}[Proof of \Cref{prop: explicit}]
To begin, we condition on the value of $X_n$ and recall the definition of the transition density $\mathsf{p}(X_{n+1}|X_n)$ from the main text:
    \begin{align}
        \E[-\log \mathsf{p}(X_{n+1}|X_n) \mid X_n] &= -\int \log(\mathsf{p}(x'|X_n)) \mathsf{p}(x'|X_n)\,\mathrm{d}\mu_{X_n}(x') \nonumber \\
        &= -r(X_n) \log r(X_n) - \int u(x'|X_n) \log u(x'|X_n) \, \mathrm{d}\lambda(x') \nonumber \\
        &= \underbrace{ -  \left(1-\int u(x'|X_n) \, \mathrm{d}\lambda(x')\right) \log \left(1-\int u(x'|X_n)\, \mathrm{d}\lambda(x')\right) }_{\text{(\ref{eq: complicated}.1)}} \nonumber \\
        & \qquad \underbrace{ - \int u(x'|X_n) \log u(x'|X_n) \, \mathrm{d}\lambda(x') }_{\text{(\ref{eq: complicated}.2)}} . \label{eq: complicated}
    \end{align}
    where we recall from the main text that $r(x)$ is the probability of rejection if the current state of the Markov chain is $x$, and $u(x'|x) = \alpha(x'|x) q(x'|x)$ is the acceptance-proposal density. 
    Thus the first term (\ref{eq: complicated}.1) in \eqref{eq: complicated} can be expressed as an expectation with respect to the proposal, as
    \begin{align*}
        \text{(\ref{eq: complicated}.1)} = -  \left(1- \mathbb{E}[ \alpha(X_{n+1}^* | X_n) | X_n] \right) \log \left( 1 - \mathbb{E}[ \alpha(X_{n+1}^* | X_n) | X_n] \right) .
    \end{align*}
    This has the form $- \mathbb{E}[Y] \log \mathbb{E}[Y]$ for $Y = 1 - \alpha(X_{n+1}^* | X_n) | X_n$, and thus from convexity of $\psi(y) = y \log y$ we may again use Jensen's inequality to obtain the lower bound $-\mathbb{E}[\psi(Y)]$, which corresponds to
    \begin{align*}
        \text{(\ref{eq: complicated}.1)} \geq - \mathbb{E}\left[ \left(1-  \alpha(X_{n+1}^* | X_n) \right) \log \left( 1 - \alpha(X_{n+1}^* | X_n)  \right) \mid X_n \right]
    \end{align*}
    The second term (\ref{eq: complicated}.2) in \eqref{eq: complicated} can be expressed as
    \begin{align*}
        \text{(\ref{eq: complicated}.2)} = - \mathbb{E}\left[ \alpha(X_{n+1}^*|X_n) [\log q(X_{n+1}^*|X_n) + \log \alpha(X_{n+1}^*|X_n)] \mid X_n \right] .
    \end{align*}
    Thus we have shown the stated inequality holds conditional on $X_n$, and taking a final expectation with respect to $X_n$ completes the argument.
\end{proof}

\section{Implementation of \ac{rmala}}
\label{ap: implement mala}

This appendix precisely describes \ac{rmala}, together with the procedure that we used to tune \ac{rmala} when the preconditioner is position-independent.

First, and for completeness, full pseudocode for \ac{rmala} is contained in \Cref{alg: rmala}.
Here $\mathrm{S}_d^+$ denotes the set of symmetric positive definite $d \times d$ matrices and $|M|$ denotes the determinant of a matrix $M \in \mathrm{S}_d^+$.
The experiments that we report in \Cref{sec: empirical} of the main text concern either the case where the preconditioner matrix $G(\cdot) = \epsilon^{-1} G_0$ is position-independent, or $G(\cdot) = \epsilon(\cdot)^{-1} G_0$ with a position-dependent scale $\epsilon(\cdot)$ learned using \ac{rlmh}.
In each case the choice of $G_0$ was as described in the main text, with sensitivity to this choice explored in \Cref{app: alternative G0}.

\begin{algorithm}[t!]
    \caption{Riemannian MALA; \citealp{girolami2011riemann}}
    \label{alg: rmala}
    \begin{algorithmic}[1]
        \Require $x_0 \in \mathbb{R}^d$ (initial state), $G(\cdot) \in \mathrm{S}_d^+$ (position-dependent preconditioner matrix), $n \in \mathbb{N}$ (number of iterations)
        \State $X_0 \gets x_0$  \Comment{initialise Markov chain}
        \For{$i = 0, 1, 2, \dots, n-1$}
            \State $X^*_{i+1} \gets \underbrace{ X_i + G(X_i)^{-1} (\nabla \log p)(X_i) }_{=: \nu(X_i)} + \sqrt{2} G(X_i)^{-1/2} Z_i$ \Comment{proposal; $Z_i \sim \mathcal{N}(0,I)$}
            \State $\displaystyle \alpha_{i} \leftarrow \min\left(1, \frac{ p(X^*_{i+1}) }{ p(X_i) }  \frac{|G(X_i)|^{1/2}}{|G(X_{i+1}^*)|^{1/2}} \frac{ \exp( - \| G(X_{i+1}^*)^{-1/2} ( X_i - \nu(X_{i+1}^*) ) \|^2 ) }{ \exp( - \| G(X_i)^{-1/2} ( X_{i+1}^* - \nu(X_i) ) \|^2 ) } \right)$ 
            \State $X_{i+1} \leftarrow X_{i+1}^*$ with probability $\alpha_{i}$, else $X_{i+1} \leftarrow X_i$ \Comment{accept/reject}
        \EndFor
        
        \State \Return Markov chain $(X_i)_{i=1}^n$
    \end{algorithmic}
\end{algorithm}

In the case of a position-independent preconditioner, we set the step size $\epsilon$ either by attempting to maximise the \ac{esjd} or tuning the \ac{aar} to 0.574, following the recommendation of \citet{roberts1998optimal}.
For tuning based on either \ac{esjd} (resp. \ac{aar}), at a fixed frequency $T$ we compare the average squared jump distance (resp. absolute deviation of the empirical acceptance rate from 0.574) $D_1$ for the chain $(X_i)_{i=n-T+1}^n$ to the corresponding quantity $D_2$ for the chain $(X_i)_{i=n-2T+1}^{n-T}$.
Based on $D_1$ and $D_2$, the step size is either increased or decreased by a multiplicative factor of 1.05 subject to remaining in an interval $[10^{-4}, 2]$; these values were chosen to ensure reasonable performance over the \texttt{posteriordb} benchmark.
In our implementation the adaptation frequency $T = 5,000$ was used.
The initial step size was set to $\epsilon = 0.1$.
Adaptation continues until such a point that adaptation is halted for the purposes of assessment, as described in the main text.

\FloatBarrier

\section{Implementation of \ac{rlmh}}
\label{ap: implement}

This appendix contains the implementational details required to reproduce the experimental results for \ac{rlmh} reported in \Cref{sec: empirical} of the main text.
First we recall the main ideas of \ac{ddpg} in \Cref{subsec: ddpg}, before reporting all training details in \Cref{subsec: training detail}.

\subsection{Deep Deterministic Policy Gradient}
\label{subsec: ddpg}

This section describes the \ac{ddpg} algorithm with reward centring for maximising $J(\theta)$ in \eqref{eq: J def}, based on the deterministic policy gradient theorem of \cite{silver2014deterministic}.
Since the algorithm itself is quite detailed, we aim simply to summarise the main aspects of \ac{ddpg} and refer the reader to \citet{lillicrap2015continuous} for full detail.

The \emph{deterministic policy gradient theorem} states that
\begin{equation}
\label{eq:dpg_theorem}
\nabla_\theta J(\pi_\theta) = \mathbb{E}_{\pi}\left[\nabla_{\theta}\pi(s) \left .\nabla_{a}\mathrm{Q}_{\pi}(s,a)  \right|_{a = \pi(s)}  \right],
\end{equation}
where the expectation here is taken with respect to the stationary distribution of the \ac{mdp} when the policy $\pi$ is fixed. 
The \emph{action-value function} $\mathrm{Q}_{\pi}(s,a)$ gives the expected discounted cumulative reward from taking an action $a$ in state $s$ and following policy $\pi$ thereafter. 

Under standard \ac{ddpg}, the policy $\pi$ and the action-value function $\mathrm{Q}$ are parameterised by neural networks whose parameters are updated via an \emph{actor-critic algorithm} based on \cite{silver2014deterministic}. Specifically, an \emph{actor network} $\pi_{\theta}(s)$ is updated in the direction of the policy gradient in \eqref{eq:dpg_theorem}, and a \emph{critic network} $Q_{\vartheta}$ that approximates the action-value function, is updated by solving the Bellman equation 
\begin{align*}
\mathrm{Q}_{\pi}(s_n, a_n) = \mathbb{E}_{\pi}\left[r_n + \gamma \mathrm{Q}_{\pi}(s_{n+1}, \pi(s_{n+1})) \right],
\end{align*}
via stochastic approximation with off-policy samples. 
In \ac{ddpg} the critic network is trained by solving the optimisation problem
\begin{align}
\label{eq:q_approx}
\argmin_{\vartheta} \mathbb{E}_{\tilde{\pi}} \left[(Q_n - \mathrm{Q}_{\vartheta}(s_n,a_n))^2\right],
\end{align}
where $Q_n = r_n + \gamma \mathrm{Q}_{\vartheta}(s_{n+1}, \pi(s_{n+1}))$ is called the \emph{one step} \ac{td} approximation, $a_n$ is generated from a stochastic \emph{behaviour policy} $\tilde{\pi}$, and $s_{n+1}$ is resulted from interacting with the environment using $a_n \sim \tilde{\pi}(s_n)$. 
The expectation in \eqref{eq:q_approx} is approximated by sampling a mini-batch from a \emph{replay buffer} that stores the experience tuples $\mathcal{D} := \{(s_i, a_i, r_i, s_{i+1})\}_{i=1}^{n}$. 
A common choice for the behaviour policy $\tilde{\pi}$ is a noisy version of the deterministic policy $\pi$, e.g., $\tilde{a}_n = a_n + \mathcal{N}_n$, where $(\mathcal{N}_n)_{n \geq 0}$ is a \emph{noise process} that must be specified. 

There are several aspects of \ac{ddpg} that are non-trivial, such as the distinction between \emph{target networks} $\mathrm{Q}_{\vartheta'}$ and $\pi_{\theta'}$ and the current actor and critic networks $\mathrm{Q}_{\vartheta}$ and $\pi_\theta$, and how these networks interact via the \emph{taming factor} $\tau$; see \citet{lillicrap2015continuous} for further detail.

Drawing on the reward‐centring framework of \citet{naik2024rewardcentering}, we replace the vanilla \ac{td} approximation $Q_n$ with the \emph{centralised} \ac{td} approximation
\begin{equation}
\bar{Q}_n = (r_n - \bar{R}) + \gamma Q_{\vartheta'} \left(s_{n+1}, \pi_{\theta}(s_{n+1})\right)
\end{equation}
where the running average of past rewards is denoted $\bar{R}$.
This centring of the immediate reward term around $\bar{R}$ has the capacity to reduce variance in the \ac{td} approximation and stabilises learning \citep{naik2024rewardcentering}.
The reward-centred variant of the \ac{ddpg} is summarised in \Cref{alg: ddpg}, and includes synchronous updates of the running average reward 
\begin{equation}
\bar{R} \leftarrow \bar{R} + \eta \cdot \frac{1}{N}\sum_{i=1}^{N} r_{(i)}
\end{equation}
where $\eta$ is a learning rate and the rewards $r_{(i)}$ are drawn from the replay buffer.

\begin{algorithm}[t!]
\begin{algorithmic}[1]
\caption{Reward-Centred Deep Deterministic Policy Gradient (DDPG)}
\label{alg: ddpg}

\Require{$s_0$ (initial state of the MDP), $\gamma \in (0,1)$ (discount factor), $\mathcal{D}$ (replay buffer), $(\mathcal{N}_t)_{t \geq 0}$ (noise process), $\alpha_\pi > 0$ (actor learning rate), $\alpha_{\mathrm{Q}} > 0$ (critic learning rate), $N$ (minibatch size),   $\eta > 0$ (centring gain), $\tau \in (0,1)$ (soft-update rate)}

\State     initialise actor parameters $\theta$ and critic parameters $\vartheta$
\State     $\pi_{\theta'} \leftarrow \pi_{\theta}$\;
    $\mathrm{Q}_{\vartheta'} \leftarrow \mathrm{Q}_{\vartheta}$ \Comment{initialise target networks}
\State    $\bar R \leftarrow 0$ \Comment{initialise average reward}

\For{each episode}
  \For{ $t=0,1,2,\dots$ until episode ends}
    \State $a_t = \pi_{\theta}(s_t) + \mathcal{N}_t$\; 
    \Comment{action with exploration noise}

    \State execute $a_t$, observe $(r_t,\,s_{t+1})$ 

    \State store transition $(s_t,a_t,r_t,s_{t+1})$ in $\mathcal{D}$

    \If{update step}
      \State sample minibatch $\{(s_i,a_i,r_i,s'_i)\}_{i=1}^N \sim \mathcal{D}$\;

      \State $r^{\text{c}}_i \leftarrow r_i - \bar R$ \; 
      \Comment{reward centring}

      \State $Q_i \leftarrow r^{\text{c}}_i + \gamma\,
                \mathrm{Q}_{\vartheta'}\bigl(s'_i,\pi_{\theta'}(s'_i)\bigr)$\;
      \Comment{TD approximation}

      \State $\vartheta \leftarrow \vartheta - \alpha_{\mathrm{Q}} \nabla_{\vartheta} \frac1N\sum_{i=1}^N\bigl(Q_i - \mathrm{Q}_{\vartheta}(s_i,a_i)\bigr)^2 $
      \Comment{critic update}

      \State $\theta \leftarrow \theta + \alpha_\pi \nabla_{\theta} \frac1N\sum_{i=1}^N \mathrm{Q}_{\vartheta}\bigl(s_i,\pi_{\theta}(s_i)\bigr)$
      \Comment{actor update}

      \State $\bar R \leftarrow \bar R + \eta\,\alpha_{\mathrm{Q}} \cdot
          \frac1N\sum_{i=1}^N (Q_i - \mathrm{Q}_{\vartheta}(s_i,a_i))$\;
      \Comment{update average reward}

      \State $\vartheta' \leftarrow (1-\tau)\,\vartheta' + \tau\,\vartheta$\;
      \Comment{soft-update critic target}

      \State $\theta' \leftarrow (1-\tau)\,\theta'+ \tau\,\theta$\;
      \Comment{soft-update actor target}
    \EndIf
  \EndFor
\EndFor
\end{algorithmic}
\end{algorithm}

For the purpose of this work we largely relied on default settings for \ac{ddpg}; full details are contained in \Cref{subsec: training detail}.
The specific design of \ac{rl} methods for use in adaptive \ac{mcmc} was not explored, but might be an interesting avenue for future work.

\subsection{Training Details}
\label{subsec: training detail}

Here we describe the settings that were used for \ac{rlmh}.
Recall that $\pi_\theta(s_n) = [\epsilon_\theta(X_n) , \epsilon_\theta(X_{n+1}^*)]$ where $s_n = [X_n , X_{n+1}^*]$ in \ac{rlmh}.

\paragraph{Parametrisation of $\epsilon_\theta$}

The function $\epsilon_\theta$ was parametrised using a fully-connected two-layer neural network with the ReLU activation function and 8 features per hidden layer; a total of $p = (8 + d) (d+1)$ parameters to be inferred.

\paragraph{Pre-training of $\epsilon_\theta$}

The parameters $\theta$ of the neural network $\epsilon_\theta$ were initialised by pre-training against the loss 
$$
\theta \mapsto \frac{1}{m} \sum_{j = 1}^m \left\| \epsilon_{\theta}(y_j) - \epsilon^\dagger \right\|^2 ,
$$ 
computed over the gold-standard samples $\{y_j\}_{j=1}^m$ with $m = 10^4$, so that the proposal corresponding to $\epsilon_\theta$ is initially a position-independent proposal with step size $\epsilon^\dagger$.
This was achieved using stochastic gradient descent \citep{robbins1951stochastic}, employing a batch size of 16, a learning rate of 0.01, and a training duration of 100 epochs.
Of course, a suitable initial step size $\epsilon^\dagger$ will be unknown in general.
Inspired by \citet{bui2012scaled}, we first considered the heuristic:
\begin{equation*}
\varepsilon_0 = \frac{\ell}{\sqrt{\lambda_{\max}(\Sigma^{-1})}, d^{1/3}} 
\end{equation*}
where $\Sigma$ is the covariance matrix computed based on gold-standard samples from the target.
This heuristic was further optimised for experiments on \texttt{posteriordb} via the nonlinear regression
\begin{equation*}
    \epsilon^\dagger = a_1 \cdot \varepsilon_0^3 + a_2 \cdot \varepsilon_0^2 + a_3 \cdot \varepsilon_0 + a_4,
\end{equation*}
with numerical coefficients determined through experimentation as $a_1 = 29$, $a_2 = -26$, $a_3 = 3.0$, and $a_4 = 1.3$.
This initialisation is merely to ensure that the difference between the initial policy $\epsilon^\dagger$ and typical learned policies remains within three orders of magnitude over the \texttt{posteriordb} benchmark, facilitating efficient and stable training on high-performance computing in our assessment of \ac{rlmh}.

\paragraph{Training of $\epsilon_\theta$}

\ac{rlmh} was performed using an implementation of \ac{ddpg} in \texttt{Python}.
The settings for training $\epsilon_\theta$ were:
\begin{itemize}
    \item 100 episodes (i.e. the number of iterations of \ac{mcmc} performed between each update based on the policy gradient), each consisting of $500$ iterations of \ac{mcmc}
    \item the noise process $(\mathcal{N}_n)_{n \geq 0}$ was Gaussian with standard deviation equal to the initial step size
    \item actor learning rate $\alpha_\pi = 10^{-6}$
    \item experience buffer length = $2.5\times 10^{4}$,
    \item minibatch size $N = 48$
    \item centering gain $\eta = 10^{-3}$
\end{itemize}

\paragraph{Parametrisation of the Critic $\mathrm{Q}$}

For all experiments we took $\mathrm{Q} : \mathbb{R}^{2d} \times \mathbb{R}^{2d} \rightarrow \mathbb{R}$ to be a fully-connected neural network with the ReLU activation function, two hidden layers, and 8 features in each hidden layer.
The parametrised critic is denoted $\mathrm{Q}_\vartheta$.

\paragraph{Training of the Critic $\mathrm{Q}_\vartheta$}

The settings for training the critic $\mathrm{Q}_\vartheta$ were identical to those for the actor, with the exceptions:
\begin{itemize}
    \item critic parameters randomly initialised by Pytorch's default initilaisation method 
    \item critic learning rate $\alpha_{\mathrm{Q}} = 10^{-2}$.
\end{itemize}

\FloatBarrier

\section{Results for the \texttt{posteriordb} Benchmark}
\label{subsec: full results}

This appendix contains additional details and results concerning the \texttt{posteriordb} benchmark.
Additional details required to reproduce our assessment are contained in \Cref{app: addit pdb}.
The \ac{mmd} performance measure is precisely defined in \Cref{app: mmd}.
The full version of \Cref{tab: posteriordb shortened} from the main text is presented in \Cref{app: pdb full}.
Sensitivity of our conclusions to the mechanism used to select $G_0$ is investigated in \Cref{app: alternative G0}.
Our investigation extended beyond \ac{rmala} and full results for the Barker proposal of \citet{livingstone2022barker} are contained and discussed in \Cref{ap: Barker}.

\subsection{Implementation Details}
\label{app: addit pdb}

The initial state $x_0 \in \mathbb{R}^d$ of all Markov chains was taken to be the arithmetic mean of $10^4$ gold-standard samples from the target.

For the experiments reported in \Cref{sec: empirical} of the main text, the static component $G_0 \in \mathrm{S}_d^+$ of the preconditioner was set to the inverse of the empirical covariance matrix of $10^4$ gold-standard samples from the target to avoid confounding due to the specific choice of how $G_0$ is estimated.
In practice, the matrix $G_0$ would also need to be estimated and this is not always straightforward; for example, one could attempt to approximate the Hessian matrix at a mode of the target, or use a simpler off-the-shelf adaptive \ac{mcmc} method in which a covariance matrix is iteratively learned \citep{andrieu2008tutorial}.

\subsection{Maximum Mean Discrepancy}
\label{app: mmd}

Performance was measured using \ac{mmd} relative to the gold-standard samples $\{y_j\}_{j=1}^m$ for each task.
Here the \ac{mmd} was based on the Gaussian kernel $k(x,y) := \exp( - \nicefrac{\|x - y\|^2}{ \ell^2} )$, whose length scale $\ell$ was determined using the \emph{median heuristic}
\begin{align*}
    \ell & := \frac{1}{2} \mathrm{median}\{ \|y_i - y_j\| : 1 \leq i , j \leq m \}
\end{align*}
following \citet{garreau2017large}.
The \ac{mmd} $D(P_n,Q_m)$ between a pair of empirical distributions $P_n = \frac{1}{n} \sum_{i=1}^n \delta_{X_i}$ and $Q_m = \frac{1}{m} \sum_{j=1}^m \delta_{y_j}$ is defined via the formula
\begin{align*}
    \mathrm{MMD}(P_n, Q_m)^2 & := \frac{1}{n^2} \sum_{i=1}^n \sum_{i' = 1}^n k(X_i,X_{i'}) - \frac{2}{nm} \sum_{i=1}^n \sum_{j=1}^m k(X_i, y_j) + \frac{1}{m^2} \sum_{j=1}^m \sum_{j'=1}^m k(y_j,y_{j'}) ,
\end{align*}
and for this work $P_n$ represents the approximation to the target $p(\cdot)$ produced using an adaptive \ac{mcmc} algorithm, and $Q_m$ represents a gold-standard set of $m = 10^4$ samples from the target, which are provided in \texttt{posteriordb}.

\subsection{Full Results for \ac{rmala}-\ac{rlmh}}
\label{app: pdb full}

Full results (i.e. the expanded version of \Cref{tab: posteriordb shortened}) are shown in \Cref{tab: pdf full}.
Instances where \ac{rmala} failed with \ac{aar} or \ac{esjd} were because proposal covariance became numerically zero; could be avoided but in principle this could introduce bias and so we simply report the failure in these cases.

\begin{table}[t!]
    \centering
    \begin{adjustbox}{max width=\textwidth}
        \begin{tabular}{|cc|c|cc|cc|}
            \hline
            &    &  \textbf{Gradient-free RLMH}  & \multicolumn{2}{|c|}{\textbf{RMALA}} & \multicolumn{2}{|c|}{\textbf{RMALA-RLMH} (proposed)} \\
                Task  & $d$  &  LESJD  & AAR  & ESJD   & LESJD  & CDLB  \\
            \hline
            \rowcolor{gray!20}
            earnings-earn\_height & 3 & 1.8(0.1)E-1& 5.2(0.3)E-1 & 5.0(0.0)E-1 & \textbf{1.4(0.1)E-2} & 1.5(0.1)E-2 \\
            \rowcolor{gray!20}
            earnings-log10earn\_height & 3 & 1.6(0.0)E-1 & 1.4(0.0)E-2 & 1.5(0.1)E-2 & 1.4(0.1)E-2 & \textbf{1.3(0.0)E-2} \\
            \rowcolor{gray!20}
            earnings-logearn\_height & 3 & 1.6(0.0)E-1 & 1.5(0.0)E-2 & 1.6(0.1)E-2 & \textbf{1.3(0.1)E-2} & 1.4(0.0)E-2 \\
            gp\_pois\_regr-gp\_regr & 3 & 1.2(0.0)E-1 & \textbf{2.2(0.1)E-2} & 2.3(0.1)E-2 & 2.4(0.1)E-2 & 2.4(0.1)E-2 \\
            \rowcolor{gray!20}
            kidiq-kidscore\_momhs & 3 & 1.5(0.0)E-1 & 1.4(0.1)E-2 & 1.5(0.1)E-2 & \textbf{1.2(0.1)E-2} & \textbf{1.2(0.0)E-2} \\
            \rowcolor{gray!20}
            kidiq-kidscore\_momiq & 3 & 1.7(0.0)E-1 & 1.5(0.1)E-2 & 1.5(0.1)E-2 & \textbf{1.2(0.0)E-2} & 1.4(0.1)E-2 \\
            \rowcolor{gray!20}
            kilpisjarvi\_mod-kilpisjarvi & 3 & 1.7(0.0)E-1 & 5.9(0.3)E-1 & 6.0(0.3)E-1 & \textbf{2.5(0.7)E-2} & \textbf{2.5(0.6)E-2} \\
            \rowcolor{gray!20}
            mesquite-logmesquite\_logvolume & 3 & 1.3(0.0)E-1 & 2.2(0.1)E-2 & 2.4(0.1)E-2 & \textbf{1.9(0.1)E-2} & 2.0(0.1)E-2 \\
            \rowcolor{gray!20}
            arma-arma11 & 4 &  1.2(0.0)E-1 & 2.3(0.1)E-2 & 2.5(0.1)E-2 & 2.1(0.1)E-2 & \textbf{2.0(0.1)E-2} \\
            \rowcolor{gray!20}
            earnings-logearn\_height\_male & 4 & 1.6(0.0)E-1 & 1.4(0.1)E-2 & 1.6(0.1)E-2 & \textbf{1.3(0.1)E-2} & \textbf{1.3(0.1)E-2} \\
            \rowcolor{gray!20}
            earnings-logearn\_logheight\_male & 4 & 1.6(0.0)E-1 & 1.3(0.1)E-2 & 1.5(0.1)E-2 & \textbf{1.1(0.1)E-2} & 1.2(0.1)E-2 \\
            garch-garch11 & 4 &  1.4(0.0)E-1 & \textbf{1.4(0.1)E-2} & 1.7(0.1)E-2 &  \cellcolor{black!50} - & 8.9(7.2)E-2 \\
            \rowcolor{gray!20}
            hmm\_example-hmm\_example & 4 &  1.3(0.0)E-1 & 2.1(0.1)E-2 & 2.2(0.1)E-2 & \textbf{1.7(0.1)E-2} & 1.9(0.1)E-2 \\
            \rowcolor{gray!20}
            kidiq-kidscore\_momhsiq & 4 &  1.4(0.0)E-1 & 1.6(0.1)E-2 & 1.8(0.1)E-2 & \textbf{1.2(0.1)E-2} & 1.7(0.1)E-2 \\
            \rowcolor{gray!20}
            earnings-logearn\_interaction & 5 & 1.4(0.0)E-1 & 1.9(0.1)E-2 & 2.1(0.1)E-2 & \textbf{1.7(0.1)E-2} & 1.8(0.1)E-2 \\
            \rowcolor{gray!20}
            earnings-logearn\_interaction\_z & 5 &  1.2(0.0)E-1 & 2.3(0.1)E-2 & 2.4(0.1)E-2 & \textbf{2.1(0.1)E-2} & \textbf{2.1(0.1)E-2} \\
            \rowcolor{gray!20}
            kidiq-kidscore\_interaction & 5 &  1.7(0.1)E-1 & 1.5(0.1)E-2 & 1.7(0.1)E-2 & \textbf{1.3(0.1)E-2} & 1.5(0.1)E-2 \\
            \rowcolor{gray!20}
            kidiq\_with\_mom\_work-kidscore\_interaction\_c & 5 & 1.6(0.0)E-1 & 1.7(0.1)E-2 & 1.8(0.1)E-2 & \textbf{1.4(0.1)E-2} & 1.5(0.1)E-2 \\
            \rowcolor{gray!20}
            kidiq\_with\_mom\_work-kidscore\_interaction\_c2 & 5 & 1.7(0.0)E-1 & 1.9(0.1)E-2 & 2.0(0.1)E-2 & \textbf{1.6(0.1)E-2} & \textbf{1.6(0.1)E-2} \\
            \rowcolor{gray!20}
            kidiq\_with\_mom\_work-kidscore\_interaction\_z & 5 & 1.5(0.1)E-1 & 2.2(0.1)E-2 & 2.3(0.1)E-2 & \textbf{1.9(0.1)E-2} & \textbf{1.9(0.1)E-2} \\
            \rowcolor{gray!20}
            kidiq\_with\_mom\_work-kidscore\_mom\_work & 5 & 1.8(0.1)E-1 & 2.3(0.1)E-2 & 2.4(0.1)E-2 & \textbf{1.9(0.1)E-2} & 2.0(0.1)E-2 \\
            \rowcolor{gray!20}
            low\_dim\_gauss\_mix-low\_dim\_gauss\_mix & 5 & 1.1(0.0)E-1 & 2.6(0.1)E-2 & 2.8(0.1)E-2 & \textbf{2.4(0.1)E-2} & 2.5(0.1)E-2 \\
            \rowcolor{gray!20}
            mesquite-logmesquite\_logva & 5 & 1.2(0.0)E-1 & 2.4(0.1)E-2 & 2.6(0.1)E-2 & \textbf{2.0(0.1)E-2} & 2.2(0.1)E-2 \\
            \rowcolor{gray!20}
            bball\_drive\_event\_0-hmm\_drive\_0 & 6 & 1.6(0.3)E-1 & 2.2(0.1)E-2 & 2.3(0.1)E-2 & \textbf{1.8(0.1)E-2} & 1.9(0.1)E-2 \\
            \rowcolor{gray!20}
            sblrc-blr & 6 & 1.7(0.0)E-1 & 1.6(0.1)E-2 & 1.7(0.1)E-2 & \textbf{1.4(0.1)E-2} & \textbf{1.4(0.1)E-2} \\
            \rowcolor{gray!20}
            sblri-blr & 6 &  1.7(0.0)E-1 & 1.6(0.1)E-2 & 1.8(0.1)E-2 & \textbf{1.4(0.1)E-2} & 1.5(0.1)E-2 \\
            \rowcolor{gray!20}
            arK-arK & 7 &  1.1(0.0)E-1 & 2.7(0.1)E-2 & 2.9(0.1)E-2 & \textbf{2.4(0.1)E-2} & \textbf{2.4(0.1)E-2} \\
            \rowcolor{gray!20}
            mesquite-logmesquite\_logvash & 7 & 1.1(0.0)E-1 & 2.5(0.1)E-2 & 2.6(0.1)E-2 & \textbf{2.1(0.1)E-2} & 2.3(0.1)E-2 \\
            \rowcolor{gray!20}
            mesquite-logmesquite & 8 &  1.1(0.0)E-1 & 3.0(0.1)E-2 & 3.2(0.1)E-2 & \textbf{2.6(0.1)E-2} & 2.8(0.1)E-2 \\
            \rowcolor{gray!20}
            mesquite-logmesquite\_logvas & 8 & 1.1(0.0)E-1 & 3.0(0.1)E-2 & 3.3(0.1)E-2 & \textbf{2.4(0.1)E-2} & 2.7(0.1)E-2 \\
            \rowcolor{gray!20}
            mesquite-mesquite & 8 &  5.1(1.1)E-1 & 3.2(0.2)E-2 & 3.4(0.1)E-2 & \textbf{2.7(0.0)E-2} & 2.9(0.1)E-2 \\
            \rowcolor{gray!20}
            eight\_schools-eight\_schools\_centered & 10 & 7.7(1.2)E-1 & 6.3(0.7)E-1 & 5.2(0.8)E-1 & \textbf{3.0(0.3)E-1} & 3.3(0.4)E-1 \\
            eight\_schools-eight\_schools\_noncentered & 10 & 1.2(0.0)E0 & 1.1(0.0)E0 & \textbf{8.1(0.0)E-1} & 8.5(0.0)E-1 & 8.5(0.0)E-1 \\
            \rowcolor{gray!20}
            nes1972-nes & 10 & 1.1(0.0)E-1 & 2.7(0.1)E-2 & 2.9(0.1)E-2 & \textbf{2.4(0.1)E-2} & 2.5(0.1)E-2 \\
            \rowcolor{gray!20}
            nes1976-nes & 10 & 1.1(0.0)E-1 & 2.9(0.1)E-2 & 3.0(0.1)E-2 & \textbf{2.6(0.1)E-2} & 2.7(0.1)E-2 \\
            \rowcolor{gray!20}
            nes1980-nes & 10 & 1.1(0.0)E-1 & 3.0(0.1)E-2 & 3.1(0.1)E-2 & \textbf{2.7(0.1)E-2} & 2.8(0.1)E-2 \\
            \rowcolor{gray!20}
            nes1984-nes & 10 & 1.2(0.0)E-1 & 3.0(0.1)E-2 & 3.1(0.1)E-2 & \textbf{2.7(0.1)E-2} & \textbf{2.7(0.1)E-2} \\
            \rowcolor{gray!20}
            nes1988-nes & 10 & 1.1(0.0)E-1 & 2.8(0.1)E-2 & 2.9(0.1)E-2 & \textbf{2.6(0.1)E-2} & \textbf{2.6(0.1)E-2} \\
            \rowcolor{gray!20}
            nes1992-nes & 10 & 1.1(0.0)E-1 & 2.8(0.1)E-2 & 2.9(0.1)E-2 & \textbf{2.6(0.1)E-2} & \textbf{2.6(0.1)E-2} \\
            \rowcolor{gray!20}
            nes1996-nes & 10 & 1.1(0.0)E-1 & 2.7(0.1)E-2 & 2.8(0.1)E-2 & \textbf{2.6(0.1)E-2} & 2.7(0.1)E-2 \\
            \rowcolor{gray!20}
            nes2000-nes & 10 & 1.2(0.0)E-1 & 2.8(0.1)E-2 & 2.8(0.1)E-2 & \textbf{2.4(0.1)E-2} & 2.5(0.1)E-2 \\
            \rowcolor{gray!20}
            gp\_pois\_regr-gp\_pois\_regr & 13 &  1.5(0.2)E0 & \cellcolor{black!50} - & \cellcolor{black!50} - & \cellcolor{black!50} - & \textbf{8.7(0.0)E-1} \\
            diamonds-diamonds & 26 &  2.0(0.1)E0 & 6.1(0.8)E-2 & \textbf{5.3(0.6)E-2} & 6.2(0.6)E-2 & 5.8(0.7)E-2 \\
            mcycle\_gp-accel\_gp & 66 &  1.9(0.0)E0 & \textbf{7.6(0.0)E-1} & 7.7(0.0)E-1 & \cellcolor{black!50} - & 7.7(0.0)E-1 \\
            \hline
        \end{tabular}
    \end{adjustbox}

\medskip
    
    \caption{Benchmarking using \texttt{posteriordb}. 
  Here we compared standard \ac{rmala} with constant step-size $\epsilon \in (0,\infty)$ tuned either by matching the \acf{aar} to 0.574 or by maximising the \acf{esjd}, to \ac{rmala} with position-dependent step size $\epsilon(\cdot)$ learned using \ac{rlmh}, based either using the \ac{lesjd} reward of \citet{wang2024reinforcement} or the proposed \ac{cdlb}. 
  The gradient-free \ac{rlmh} method of \citet{wang2024reinforcement} is also included as a benchmark.
  Performance was measured using \acf{mmd} based on the Gaussian kernel with lengthscale selected using the median heuristic applied to $10,000$ gold-standard samples from the target, and $d$ is the dimension of the target. 
  Results are based on an average of 10 replicates, with standard errors (in parentheses) reported.
  The method with smallest average \ac{mmd} is highlighted in \textbf{bold}, gray rows indicate tasks for which \ac{rlmh} performed best, and dashes indicate where catastrophic failure occurred during training of \ac{rlmh}.}
  \label{tab: pdf full}
\end{table}

\subsection{Exploring the Sensitivity to $G_0$}
\label{app: alternative G0}

The results that we report for \texttt{posteriordb} in the main text set $G_0$ based on $10^4$ gold-standard samples from the target.
However, this is an idealised situation since one will not in general have such details information on the target at the outset of running \ac{mcmc}.
Practical strategies include attempting to approximate the Hessian matrix at a mode of the target, or using a simpler off-the-shelf adaptive \ac{mcmc} method in which a covariance matrix is iteratively learned \citep{andrieu2008tutorial}.
This appendix investigates the performance of \ac{rmala} when $G_0$ is also to be estimated.
To this end, we reproduced the experiment from \Cref{tab: posteriordb shortened} in the main text but with $G_0$ set equal to the negative Hessian of $\log p$ evaluated at the initial state $x_0$ of the Markov chain.
Full results are displayed in \Cref{tab: hess full}.
In summary, the broad conclusions that we draw in the main text, regarding the potential to improve performance using a position-dependent step size $\epsilon(\cdot)$ trained using \ac{rlmh} and the \ac{cdlb} reward, continue to hold, indicating insensitivity of these conclusions to the precise mechanism used to select $G_0$.

\begin{table}[t!]
    \centering
    \begin{adjustbox}{max width=0.8\textwidth}
        \begin{tabular}{|cc|cc|cc|}
            \hline
            &    & \multicolumn{2}{|c|}{\textbf{RMALA}} & \multicolumn{2}{|c|}{\textbf{RMALA-RLMH} (proposed)} \\
            Task  & $d$  & AAR  & ESJD   & LESJD  & CDLB  \\
            \hline
            \rowcolor{gray!20}
            earnings-earn\_height & 3 & 5.2(0.3)E-1 & 1.9(0.2)E-2 & 1.5(0.1)E-2 & \textbf{1.4(0.2)E-2} \\
            \rowcolor{gray!20}
            earnings-log10earn\_height & 3 & 1.4(0.0)E-2 & 1.6(0.1)E-2 & 1.4(0.1)E-2 & \textbf{1.3(0.1)E-2} \\
            \rowcolor{gray!20}
            earnings-logearn\_height & 3 & 1.5(0.0)E-2 & 1.6(0.1)E-2 & \textbf{1.4(0.1)E-2} & \textbf{1.4(0.1)E-2} \\
            gp\_pois\_regr-gp\_regr & 3 & \textbf{2.2(0.1)E-2} & 2.3(0.1)E-2 & 2.4(0.1)E-2 & 2.4(0.1)E-2 \\
            \rowcolor{gray!20}
            kidiq-kidscore\_momhs & 3 & 1.4(0.1)E-2 & 1.5(0.1)E-2 & \textbf{1.2(0.1)E-2} & \textbf{1.2(0.1)E-2} \\
            \rowcolor{gray!20}
            kidiq-kidscore\_momiq & 3 & 1.4(0.0)E-2 & 1.6(0.1)E-2 & \textbf{1.1(0.1)E-2} & 1.4(0.1)E-2 \\
            kilpisjarvi\_mod-kilpisjarvi & 3 & 5.9(0.3)E-1 & \textbf{2.0(0.4)E-2} & 2.5(0.7)E-2 & 2.5(0.7)E-2 \\
            \rowcolor{gray!20}
            mesquite-logmesquite\_logvolume & 3 & 2.3(0.1)E-2 & 2.4(0.1)E-2 & \textbf{1.9(0.1)E-2} & 2.0(0.1)E-2 \\
            \rowcolor{gray!20}
            arma-arma11 & 4 & 2.3(0.1)E-2 & 2.5(0.1)E-2 & \textbf{2.1(0.1)E-2} & \textbf{2.1(0.1)E-2} \\
            \rowcolor{gray!20}
            earnings-logearn\_height\_male & 4 & 1.4(0.1)E-2 & 1.5(0.1)E-2 & \textbf{1.2(0.1)E-2} & 1.3(0.0)E-2 \\
            \rowcolor{gray!20}
            earnings-logearn\_logheight\_male & 4 & 1.3(0.1)E-2 & 1.4(0.0)E-2 & \textbf{1.1(0.1)E-2} & 1.2(0.1)E-2 \\
            garch-garch11 & 4 & \textbf{1.8(0.1)E-2} & 2.0(0.1)E-2 & \cellcolor{black!50} - & 1.9(0.2)E-2 \\
            \rowcolor{gray!20}
            hmm\_example-hmm\_example & 4 & 2.1(0.1)E-2 & 2.2(0.1)E-2 & \textbf{1.8(0.1)E-2} & 1.9(0.1)E-2 \\
            \rowcolor{gray!20}
            kidiq-kidscore\_momhsiq & 4 & 1.6(0.1)E-2 & 1.8(0.1)E-2 & \textbf{1.3(0.1)E-2} & 1.7(0.1)E-2 \\
            \rowcolor{gray!20}
            one\_comp\_mm\_elim\_abs-one\_comp\_mm\_elim\_abs & 4 & 2.4(0.1)E-2 & 2.6(0.1)E-2 & \textbf{1.2(0.8)E-1} & 2.0(0.2)E-2 \\
            \rowcolor{gray!20}
            earnings-logearn\_interaction & 5 & 1.9(0.1)E-2 & 2.0(0.1)E-2 & \textbf{1.7(0.1)E-2} & \textbf{1.7(0.1)E-2} \\
            \rowcolor{gray!20}
            earnings-logearn\_interaction\_z & 5 & 2.3(0.1)E-2 & 2.4(0.1)E-2 & \textbf{2.1(0.1)E-2} & \textbf{2.1(0.1)E-2} \\
            \rowcolor{gray!20}
            kidiq-kidscore\_interaction & 5 & 1.5(0.1)E-2 & 1.7(0.1)E-2 & \textbf{1.3(0.1)E-2} & 1.5(0.1)E-2 \\
            \rowcolor{gray!20}
            kidiq\_with\_mom\_work-kidscore\_interaction\_c & 5 & 1.7(0.1)E-2 & 1.8(0.1)E-2 & \textbf{1.4(0.1)E-2} & 1.5(0.1)E-2 \\
            \rowcolor{gray!20}
            kidiq\_with\_mom\_work-kidscore\_interaction\_c2 & 5 & 1.9(0.1)E-2 & 2.0(0.1)E-2 & \textbf{1.6(0.1)E-2} & \textbf{1.6(0.1)E-2} \\
            \rowcolor{gray!20}
            kidiq\_with\_mom\_work-kidscore\_interaction\_z & 5 & 2.2(0.1)E-2 & 2.3(0.1)E-2 & \textbf{1.9(0.1)E-2} & \textbf{1.9(0.1)E-2} \\
            \rowcolor{gray!20}
            kidiq\_with\_mom\_work-kidscore\_mom\_work & 5 & 2.3(0.1)E-2 & 2.4(0.1)E-2 & \textbf{1.9(0.1)E-2} & 2.0(0.1)E-2 \\
            \rowcolor{gray!20}
            low\_dim\_gauss\_mix-low\_dim\_gauss\_mix & 5 & 2.7(0.1)E-2 & 2.8(0.1)E-2 & \textbf{2.4(0.1)E-2} & \textbf{2.4(0.1)E-2} \\
            \rowcolor{gray!20}
            mesquite-logmesquite\_logva & 5 & 2.4(0.1)E-2 & 2.6(0.1)E-2 & \textbf{2.1(0.1)E-2} & 2.2(0.1)E-2 \\
            \rowcolor{gray!20}
            bball\_drive\_event\_0-hmm\_drive\_0 & 6 & 2.3(0.1)E-2 & 2.4(0.1)E-2 & \textbf{1.9(0.1)E-2} & 2.0(0.1)E-2 \\
            \rowcolor{gray!20}
            bball\_drive\_event\_1-hmm\_drive\_1 & 6 & 1.9(0.1)E-2 & 2.0(0.1)E-2 & \textbf{1.6(0.1)E-2} & 1.9(0.1)E-2 \\
            \rowcolor{gray!20}
            sblrc-blr & 6 & 1.5(0.1)E-2 & 1.8(0.2)E-2 & \textbf{1.4(0.1)E-2} & \textbf{1.4(0.1)E-2} \\
            \rowcolor{gray!20}
            sblri-blr & 6 & 1.6(0.1)E-2 & 1.7(0.1)E-2 & \textbf{1.4(0.1)E-2} & \textbf{1.4(0.1)E-2} \\
            \rowcolor{gray!20}
            arK-arK & 7 & 2.7(0.1)E-2 & 2.9(0.1)E-2 & \textbf{2.4(0.1)E-2} & 2.5(0.1)E-2 \\
            \rowcolor{gray!20}
            mesquite-logmesquite\_logvash & 7 & 2.5(0.1)E-2 & 2.7(0.1)E-2 & \textbf{2.2(0.1)E-2} & 2.3(0.1)E-2 \\
            \rowcolor{gray!20}
            hudson\_lynx\_hare-lotka\_volterra & 8 & 3.0(0.1)E-2 & 3.4(0.1)E-2 & \textbf{2.3(0.1)E-2} & 2.8(0.1)E-2 \\
            \rowcolor{gray!20}
            mesquite-logmesquite & 8 & 3.1(0.1)E-2 & 3.3(0.1)E-2 & \textbf{2.7(0.1)E-2} & 3.0(0.1)E-2 \\
            \rowcolor{gray!20}
            mesquite-logmesquite\_logvas & 8 & 3.1(0.1)E-2 & 3.3(0.1)E-2 & \textbf{2.6(0.2)E-2} & 3.0(0.1)E-2 \\
            \rowcolor{gray!20}
            mesquite-mesquite & 8 & 3.3(0.2)E-2 & 3.4(0.2)E-2 & 3.0(0.0)E-2 & \textbf{2.9(0.2)E-2} \\
            eight\_schools-eight\_schools\_centered & 10 & 6.3(0.7)E-1 & \textbf{5.2(0.8)E-1} & 5.7(0.4)E-1 & 7.7(0.6)E-1 \\
            eight\_schools-eight\_schools\_noncentered & 10 & \textbf{6.2(0.0)E-1} & 6.3(0.0)E-1 & 6.3(0.0)E-1 & 6.3(0.0)E-1 \\
            \rowcolor{gray!20}
            nes1972-nes & 10 & 2.7(0.1)E-2 & 2.8(0.1)E-2 & \textbf{2.4(0.1)E-2} & 2.5(0.1)E-2 \\
            \rowcolor{gray!20}
            nes1976-nes & 10 & 2.9(0.1)E-2 & 2.9(0.1)E-2 & \textbf{2.5(0.1)E-2} & 2.6(0.1)E-2 \\
            \rowcolor{gray!20}
            nes1980-nes & 10 & 3.1(0.1)E-2 & 3.1(0.1)E-2 & \textbf{2.7(0.1)E-2} & 2.8(0.1)E-2 \\
            \rowcolor{gray!20}
            nes1984-nes & 10 & 3.0(0.1)E-2 & 3.0(0.1)E-2 & \textbf{2.7(0.1)E-2} & \textbf{2.7(0.1)E-2} \\
            \rowcolor{gray!20}
            nes1988-nes & 10 & 2.8(0.1)E-2 & 2.8(0.1)E-2 & \textbf{2.5(0.1)E-2} & \textbf{2.5(0.1)E-2} \\
            \rowcolor{gray!20}
            nes1992-nes & 10 & 2.8(0.1)E-2 & 2.9(0.1)E-2 & \textbf{2.5(0.1)E-2} & 2.6(0.1)E-2 \\
            \rowcolor{gray!20}
            nes1996-nes & 10 & 2.8(0.1)E-2 & 2.8(0.1)E-2 & \textbf{2.5(0.1)E-2} & 2.6(0.1)E-2 \\
            \rowcolor{gray!20}
            nes2000-nes & 10 & 2.8(0.1)E-2 & 2.9(0.1)E-2 & \textbf{2.4(0.1)E-2} & \textbf{2.4(0.1)E-2} \\
            \rowcolor{gray!20}
            gp\_pois\_regr-gp\_pois\_regr & 13 & \cellcolor{black!50} - & \cellcolor{black!50} - & \textbf{1.1(0.0)E0} & 8.3(1.1)E-1 \\
            diamonds-diamonds & 26 & 5.3(0.6)E-2 & \textbf{4.7(0.4)E-2} & 5.8(0.6)E-2 & 5.6(0.5)E-2 \\
            mcycle\_gp-accel\_gp & 66 & \textbf{7.6(0.0)E-1} & 7.7(0.0)E-1 & \cellcolor{black!50} - & 8.2(0.5)E-1 \\
            \hline
        \end{tabular}
    \end{adjustbox}

\medskip
    
    \caption{Benchmarking using \texttt{posteriordb}. 
  Here we compared standard \ac{rmala} with constant step-size $\epsilon \in (0,\infty)$ tuned either by matching the \acf{aar} to 0.574 or by maximising the \acf{esjd}, to \ac{rmala} with position-dependent step size $\epsilon(\cdot)$ learned using \ac{rlmh}, based either using the \ac{lesjd} reward of \citet{wang2024reinforcement} or the proposed \ac{cdlb}. 
  In contrast to the experimental study reported in the main text, here the matrix $G_0$ was selected as the negative Hessian matrix at the initial state of the Markov chain. 
  Performance was measured using \acf{mmd} based on the Gaussian kernel with lengthscale selected using the median heuristic applied to $10,000$ gold-standard samples from the target, and $d$ is the dimension of the target. 
  Results are based on an average of 10 replicates, with standard errors (in parentheses) reported.
  The method with smallest average \ac{mmd} is highlighted in \textbf{bold}, gray rows indicate tasks for which \ac{rlmh} performed best, and dashes indicate where catastrophic failure occurred during training of \ac{rlmh}.}
  \label{tab: hess full}
\end{table}

\subsection{Additional Results for the Barker Proposal}
\label{ap: Barker}

The \emph{Barker proposal} was developed in \citet{livingstone2022barker} as an alternative to \ac{mala} that trades efficiency for being more easily tuned.
It is based on the accept-reject rule of \citet{barker1965monte}, stated for $d$-dimensional distributions in \Cref{alg:barkerMH 1}.

To investigate the potential of \ac{rlmh} in the setting of the Barker proposal, we consider allowing the proposal variance $\sigma$ in \Cref{alg:barkerMH 1} to be state-dependent.
That is, the parameter $\sigma \equiv \sigma(x)$ will depend on the current state $x$ of the Markov chain; we set $\sigma(x) = \sigma_\theta(x)$ where $\sigma_\theta : \mathbb{R}^d \rightarrow \mathbb{R}$ is a neural network to be trained using \ac{rl}.
That is, we have an \ac{mdp} with action of the form $a_n = [\sigma_\theta(X_n), \sigma_\theta(X_{n+1}^*)]$.
The full formulation of the Metropolis--Hastings algorithm based on the state-dependent Barker proposal is clarified in \Cref{alg:barkerMH}.

For \ac{rlmh} we employed the same settings as used for \ac{rmala}; no fine-tuning was performed.
As a baseline we simply considered the constant step size $\epsilon = \epsilon^\dagger$ described in \Cref{subsec: training detail}.
Results in \Cref{tab: barker mean} indicate little benefit from \ac{rlmh} in this setting; this makes sense and is consistent with the goal of the Barker proposal to be robust to how the step size is selected.
However, our results do not rule out a benefit from using \ac{rlmh} in the context of the Barker proposal, since we did not fine tune settings for \ac{rlmh}.

\begin{algorithm}[t!]
    \caption{Barker proposal on $\R^d$; Algorithm 1 of \citealp{livingstone2022barker}.}
    \label{alg:barkerMH 1}
    \begin{algorithmic}[1]
    \Require $x \in \R^d$ (current state), scale $\sigma >0$ (scale).  
    \For{$i \in \{1 , \dots , d \}$}
        \State Draw $Z_i \sim \mathcal{N}(0,\sigma^2)$.
        \State Set $b_i = 1$ with probability
        $$
        \frac{1}{1 + e^{- Z_i \partial_i \log p(x)}}
        $$
        \hspace{11pt} else set $b_i = -1$.
        \State Set $X_i^* = x_i + b_i Z_i$.
    \EndFor
    \State \Return proposed state $X^* \in \mathbb{R}^d$.
    \end{algorithmic}
\end{algorithm}

\begin{algorithm}[t!]
    \caption{Metropolis--Hastings with the state-dependent Barker proposal on $\R^d$. [Here $\upphi(x)$ is the probability density function of the standard normal distribution on $\mathbb{R}$.]}
    \label{alg:barkerMH}
    \begin{algorithmic}[1]
    \Require $x_0 \in \R^d$ (initial state), $\sigma(\cdot) > 0$ (and state-dependent scale), $n \in \mathbb{N}$ (number of iterations).  
    \State $X_0 \gets x_0$ \Comment{initialise Markov chain}
    \For{$t = 0,1,2,\dots,N-1$}
        \State Given the current state $X_i$, sample $X_{i+1}^*$ using \Cref{alg:barkerMH 1} with scale $\sigma(X_i)$.
        \State Set $X_{i+1} \gets X_{i+1}^*$ with probability $$
            \min\left(  1,    \frac{p(X_{i+1}^*)}{p(X_i)} \times  \prod_{j=1}^d               \left.   \frac{\frac{1}{\sigma(X_{i+1}^*)}\upphi\left(\frac{X_{i,j}-X_{i+1,j}^*}{\sigma(X_{i+1}^*)} \right)}{1+e^{(X_{i+1,j}^*-X_{i,j})\partial_j\log p(X_{i+1}^*)}}    \middle/                 \frac{\frac{1}{\sigma(X_i)}\upphi\left(\frac{X_{i+1,j}^*-X_{i,j}}{\sigma(X_i)} \right)}{1+e^{(X_{i,j}-X_{i+1,j}^*)\partial_j \log p(X_i)}}               \right.           \right) $$ \hspace{11pt} else set set $X_{i+1} \gets X_i$
        \EndFor
        \State \Return Markov chain $\{X_0,\dots,X_n\}$.
        \end{algorithmic}
\end{algorithm}

\begin{table}[t!]
  \centering
  \begin{adjustbox}{max width=0.7\textwidth}
    \begin{tabular}{|cc|c|cc|}
      \hline
      \multicolumn{2}{|c|}{} & \multicolumn{1}{c|}{\textbf{Barker}} & \multicolumn{2}{c|}{\textbf{Barker-RLMH}} \\
      Task & $d$ & Constant  & LESJD  & CDLB \\
            \hline
earnings-earn\_height & 3 & 5.0(0.0)E-1 & 5.0(0.0)E-1 & 5.0(0.0)E-1 \\
earnings-log10earn\_height & 3 & 5.5(0.0)E-1 & 5.5(0.0)E-1 & 5.5(0.0)E-1 \\
earnings-logearn\_height & 3 & 1.1(0.0)E0 & 1.1(0.0)E0 & 1.1(0.0)E0 \\
\rowcolor{gray!20}
gp\_pois\_regr-gp\_regr & 3 & 1.1(0.1)E-2 & \textbf{2.5(0.3)E-3} & 2.6(0.2)E-3 \\
\rowcolor{gray!20}
kidiq-kidscore\_momhs & 3 & 4.0(0.8)E-2 & \textbf{3.6(1.4)E-2} & 3.9(0.9)E-2 \\
\rowcolor{gray!20}
kidiq-kidscore\_momiq & 3 & 4.8(0.2)E-1 & \textbf{4.6(0.2)E-1} & 4.7(0.2)E-1 \\
kilpisjarvi\_mod-kilpisjarvi & 3 & 6.3(0.0)E-1 & 6.3(0.0)E-1 & 6.3(0.0)E-1 \\
\rowcolor{gray!20}
mesquite-logmesquite\_logvolume & 3 & 4.9(0.6)E-1 & \textbf{3.2(0.7)E-1} & 3.7(0.6)E-1 \\
arma-arma11 & 4 & 10.0(0.0)E-1 & 10.0(0.0)E-1 & 10.0(0.0)E-1 \\
earnings-logearn\_height\_male & 4 & 4.9(0.0)E-1 & 4.9(0.0)E-1 & 4.9(0.0)E-1 \\
earnings-logearn\_logheight\_male & 4 & 6.5(0.0)E-1 & 6.5(0.0)E-1 & 6.5(0.0)E-1 \\
\rowcolor{gray!20}
garch-garch11 & 4 & 3.7(0.5)E-2 & \textbf{1.6(0.2)E-2} & 2.1(0.5)E-2 \\
\rowcolor{gray!20}
hmm\_example-hmm\_example & 4 & 5.6(0.9)E-1 & \textbf{3.1(0.6)E-1} & 4.2(0.7)E-1 \\
kidiq-kidscore\_momhsiq & 4 & 1.2(0.0)E0 & 1.2(0.0)E0 & 1.2(0.0)E0 \\
one\_comp\_mm\_elim\_abs-one\_comp\_mm\_elim\_abs & 4 & \textbf{1.1(0.4)E-1} & 7.4(0.0)E-1 & 7.4(0.0)E-1 \\
earnings-logearn\_interaction & 5 & 6.7(0.0)E-1 & 6.7(0.0)E-1 & 6.7(0.0)E-1 \\
earnings-logearn\_interaction\_z & 5 & 6.7(0.0)E-1 & 6.7(0.0)E-1 & 6.7(0.0)E-1 \\
kidiq-kidscore\_interaction & 5 & 1.0(0.0)E0 & 1.0(0.0)E0 & 1.0(0.0)E0 \\
\rowcolor{gray!20}
kidiq\_with\_mom\_work-kidscore\_interaction\_c & 5 & 9.1(0.8)E-1 & \textbf{6.0(0.7)E-1} & 6.7(0.7)E-1 \\
\rowcolor{gray!20}
kidiq\_with\_mom\_work-kidscore\_interaction\_c2 & 5 & 7.8(0.6)E-1 & \textbf{5.8(0.7)E-1} & \textbf{5.8(0.7)E-1} \\
kidiq\_with\_mom\_work-kidscore\_interaction\_z & 5 & \textbf{3.6(0.3)E-2} & 4.2(1.1)E-2 & 3.8(0.8)E-2 \\
\rowcolor{gray!20}
kidiq\_with\_mom\_work-kidscore\_mom\_work & 5 & 1.2(0.3)E-1 & \textbf{7.4(1.7)E-2} & 7.4(1.7)E-2 \\
low\_dim\_gauss\_mix-low\_dim\_gauss\_mix & 5 & 6.9(0.0)E-1 & 6.9(0.0)E-1 & 6.9(0.0)E-1 \\
mesquite-logmesquite\_logva & 5 & 6.3(0.0)E-1 & 6.3(0.0)E-1 & 6.3(0.0)E-1 \\
\rowcolor{gray!20}
bball\_drive\_event\_0-hmm\_drive\_0 & 6 & 5.9(0.0)E-1 & \textbf{5.1(0.4)E-1} & 5.4(0.3)E-1 \\
bball\_drive\_event\_1-hmm\_drive\_1 & 6 & 1.1(0.0)E0 & 1.1(0.0)E0 & 1.1(0.0)E0 \\
sblrc-blr & 6 & 6.1(0.0)E-1 & 6.1(0.0)E-1 & 6.1(0.0)E-1 \\
sblri-blr & 6 & 5.4(0.0)E-1 & 5.4(0.0)E-1 & 5.4(0.0)E-1 \\
arK-arK & 7 & 7.2(0.0)E-1 & 7.2(0.0)E-1 & 7.2(0.0)E-1 \\
mesquite-logmesquite\_logvash & 7 & 8.8(0.0)E-1 & 8.8(0.0)E-1 & 8.8(0.0)E-1 \\
hudson\_lynx\_hare-lotka\_volterra & 8 & 9.2(0.0)E-1 & 9.2(0.0)E-1 & 9.2(0.0)E-1 \\
mesquite-logmesquite & 8 & 8.7(0.0)E-1 & 8.7(0.0)E-1 & 8.7(0.0)E-1 \\
mesquite-logmesquite\_logvas & 8 & 7.3(0.0)E-1 & 7.3(0.0)E-1 & 7.3(0.0)E-1 \\
mesquite-mesquite & 8 & 8.1(0.1)E-1 & 8.1(0.1)E-1 & 8.1(0.1)E-1 \\
eight\_schools-eight\_schools\_centered & 10 & \textbf{4.3(0.5)E-2} & 5.4(0.3)E-1 & 5.4(0.3)E-1 \\
\rowcolor{gray!20}
eight\_schools-eight\_schools\_noncentered & 10 & 5.7(0.1)E-1 & \textbf{5.6(0.0)E-1} & 5.7(0.0)E-1 \\
nes1972-nes & 10 & 8.4(0.0)E-1 & 8.4(0.0)E-1 & 8.4(0.0)E-1 \\
nes1976-nes & 10 & 7.1(0.0)E-1 & 7.1(0.0)E-1 & 7.1(0.0)E-1 \\
nes1980-nes & 10 & 6.4(0.0)E-1 & 6.4(0.0)E-1 & 6.4(0.0)E-1 \\
nes1984-nes & 10 & 1.1(0.0)E0 & 1.1(0.0)E0 & 1.1(0.0)E0 \\
nes1988-nes & 10 & 6.6(0.0)E-1 & 6.6(0.0)E-1 & 6.6(0.0)E-1 \\
nes1992-nes & 10 & 1.2(0.0)E0 & 1.2(0.0)E0 & 1.2(0.0)E0 \\
nes1996-nes & 10 & 7.2(0.0)E-1 & 7.2(0.0)E-1 & 7.2(0.0)E-1 \\
nes2000-nes & 10 & 1.1(0.0)E0 & 1.1(0.0)E0 & 1.1(0.0)E0 \\
gp\_pois\_regr-gp\_pois\_regr & 13 & \textbf{8.4(0.6)E-1} & 8.7(0.0)E-1 & 8.7(0.0)E-1 \\
diamonds-diamonds & 26 & 6.8(0.0)E-1 & 6.8(0.0)E-1 & 6.8(0.0)E-1 \\
mcycle\_gp-accel\_gp & 66 & 7.7(0.0)E-1 & 7.7(0.0)E-1 & 7.7(0.0)E-1 \\
\hline
\end{tabular}
    \end{adjustbox}

\medskip
    
    \caption{Benchmarking using \texttt{posteriordb}. 
  Here we compared standard Barker \ac{mcmc} with constant step-size $\epsilon$ to a position-dependent step size $\epsilon(\cdot)$ learned using \ac{rlmh}, based either using the \ac{lesjd} reward of \citet{wang2024reinforcement} or the proposed \ac{cdlb}. 
  Performance was measured using \acf{mmd} based on the Gaussian kernel with lengthscale selected using the median heuristic applied to $10,000$ gold-standard samples from the target, and $d$ is the dimension of the target. 
  Results are based on an average of 10 replicates, with standard errors (in parentheses) reported.
  The method with smallest average \ac{mmd} is highlighted in \textbf{bold}, gray rows indicate tasks for which \ac{rlmh} performed best.}
  \label{tab: barker mean}
\end{table}

\end{document}